\documentclass[preprint,aps,amsmath,amssymb,nofootinbib,12pt]{revtex4}

\usepackage{epsfig}
\usepackage{slashed}
\usepackage{graphicx}
\usepackage{multirow,color}
\usepackage{amsmath}
\usepackage{float}
\usepackage{diagbox}
\usepackage{CJK}
\usepackage{color}
\usepackage{xcolor}
\usepackage{times}
\usepackage{subfigure}
\usepackage{bm}
\usepackage{braket}
\usepackage{booktabs}
\usepackage{array}
\usepackage[mathscr]{euscript}
\usepackage{caption}
\usepackage{makecell}
\usepackage{epstopdf}
\usepackage{hyperref}
\usepackage{subcaption}

\makeatletter

\newcommand{\Rmnum}[1]{\expandafter\@slowromancap\romannumeral #1@}
\makeatother
\graphicspath{{fig/}}
\begin{document}

\title{Probing Neutral Triple Gauge Couplings at $e^{-}p$ colliders}
	
\author{Xue-Jia Cheng$^{1,2}$}
\thanks{xjcheng@lnnu.edu.cn}
\author{Chong-Xing Yue$^{1,2}$}
\thanks{cxyue@lnnu.edu.cn}
\author{Ji-Chong Yang$^{1,2}$}
\thanks{yangjichong@lnnu.edu.cn}
\author{Si-Tong Liu$^{1,2}$}
\thanks{13634669801@163.com}
	
\date{\today}
\affiliation{
$^1$Department of Physics, Liaoning Normal University, Dalian 116029, China\\
$^2$Center for Theoretical and Experimental High Energy Physics, Liaoning Normal University, Dalian 116029, China
}

\begin{abstract}	
The Standard Model Effective Field Theory~(SMEFT) has attracted much attention as a model-independent way for probing new physical signals. 
In the SMEFT framework, $e^{-}p$ colliders can be used to study new interactions of gauge bosons, such as neutral Triple Gauge Couplings~(nTGCs). 
We study the signals and backgrounds of six different processes at the Future Circular Collider-hadron electron~(FCC-he) and propose different event selection strategies. 
The expected constraints on the coefficients of the dimension-8 operators for each process, as well as the combined constraints are obtained. 
The results show that the sensitivity of $e^{-}p$ colliders to nTGCs is similar to the current Large Hadron Collider~(LHC) experiments and is more competitive than that of Circular Electron Positron Collider~(CEPC) experiments.
\end{abstract}
	
\maketitle

\section{\label{level1} Introduction }
Searching for New Physics (NP) beyond the Standard Model~(SM) is one of the main goals of current high-energy physics. 
The SM Effective Field Theory (SMEFT) ~\cite{Weinberg:1979sa,Weinberg:2021exr,Grzadkowski:2010es,Willenbrock:2014bja,Masso:2014xra} provides a general framework for converting NP emergent low-energy effects parameterized as SM field-effective operators. 
If the results of collider experiments deviate from the SM predictions, then the SMEFT can be used to describe this deviation and make predictions in a model-independent way. 
In the absence of significant deviations, the experimental results can effectively constraints the coefficients in the SMEFT, which in turn can be translated into constraints on the masses and coupling strengths of the new particles in the different NP models.

While current research on the SMEFT remains centered on dimension-6 operators~\cite{Ge:2016zro,Ellis:2018gqa,LHCHiggsCrossSectionWorkingGroup:2016ypw,Ellis:2020unq}, the phenomenological landscape has recently undergone a paradigm shift with dimension-8 operators emerging as a frontier in both theoreticals and experimental investigation~\cite{Ellis:2017edi,Ellis:2018cos,Alday:2014qfa}.    
For example, neutral Triple Gauge Couplings~(nTGCs) ~\cite{Gounaris:2000dn,Degrande:2013kka,Senol:2018cks,Gago:2001si,Ellis:2025ghl,Semushin:2025uts,Ellis:2019zex,Ellis:2020ljj,Fu:2021mub,Ellis:2022zdw,Gounaris:1999kf} provide a unique way to explore NP beyond the SM and are attracting increasing theoretical and experimental interest. 
It is known that they do not appear in the SM Lagrangian, nor do they arise from the dimension-6 operators in the SMEFT, which indicates that nTGCs emerges starting at dimension-8. 
That is, any signs of a non-vanishing nTGCs would be direct evidence of new physics and nTGCs are powerful probes to detect NP indirect effects beyond the SM~\cite{Ellis:2023ucy,Ellis:2019zex,Semushin:2025uts,Fu:2021mub,Ellis:2020ljj,ATLAS:2018nci,CMS:2020gtj,ATLAS:2025ply,Rahaman:2018ujg,Gutierrez-Rodriguez:2023rab,Guo:2024qyx,Ellis:2022zdw,Ananthanarayan:2014sea,Rahaman:2016pqj,Rahaman:2017qql,Yang:2021kyy,Ananthanarayan:2011fr,Spor:2022zob,Liu:2024tcz,Subba:2023jia,Jahedi:2022duc,Jahedi:2023myu,Gounaris:1999kf,Ellis:2025jgt}.
Previous studies have thoroughly studied probing nTGCs at $pp$ colliders~\cite{Gounaris:1999kf,Semushin:2025uts,Ellis:2022zdw,ATLAS:2018nci,CMS:2020gtj,ATLAS:2025ply,Ellis:2023ucy,Rahaman:2018ujg} and $e^+e^-$ colliders~\cite{Ananthanarayan:2014sea,Rahaman:2016pqj,Rahaman:2017qql,Ellis:2019zex,Ellis:2020ljj,Yang:2021kyy,Ananthanarayan:2011fr,Fu:2021mub,Spor:2022zob,Liu:2024tcz,Subba:2023jia,Jahedi:2022duc,Jahedi:2023myu,Guo:2024qyx,Gutierrez-Rodriguez:2023rab,Ellis:2025jgt}, while the study of dimension-8 operators for nTGCs at $e^-p$ colliders needs to be continued and refined. This study aims to comprehensively understand the estimation of  nTGCs in future colliders.

The $e^-p$ colliders have great physical potential and provide one of the new windows to NP. Compared to $pp$ colliders and  $e^+e^-$ colliders, $e^-p$ colliders combine their excellent properties to provide not only cleaner backgrounds but also suppression of backgrounds from strong QCD interactions, and the ability of $e^-p$ colliders to study deep inelastic lepton scattering from the internal structure of nuclei~\cite{LHeC:2020van, LHeCStudyGroup:2012zhm}.
In addition, since the initial states are asymmetric, the backward and forward scattering can be disentangled, which can greatly improve the significance of the signal~\cite{Spor:2021dhf}. The $e^{-}p$ colliders are able to radiate gauge bosons from the electron and proton beam, respectively, making it possible to measure the interactions between gauge bosons~\cite{Kuze:2018dqd}. Therefore, $e^-p$ colliders are not only a good choice to complement $pp$ and $e^+e^-$ colliders, but also provide a unique way to accurately study Higgs physics, top quarks, electroweak physics, and new physics beyond the SM~\cite{Hesari:2018ssq}.

In this work, we will base our analysis on the $e^-p$ colliders such as Large Hadron electron Collider (LHeC)~\cite{LHeC:2020van, LHeCStudyGroup:2012zhm, Ahmadova:2025vzd} and Future Circular Collider-hadron electron~(FCC-he)~\cite{FCC:2018byv}. The LHeC is expected to operate at the center-of-mass energy $\sqrt{s}=1.30$ TeV with the integrated luminosity $\mathcal{L} = 1\;\rm{ab}^{-1}$, and electron and proton beam energies at $60$ GeV and $7$ TeV.  The FCC-he is expected to operate at $\sqrt{s}=3.50\;(5.29)$ TeV and $\mathcal{L} =1~(2)\;\rm{ab}^{-1}$, and corresponding electron and proton beam energies at $60\;(140)$ GeV and $50$ TeV, respectively.
Using Monte Carlo~(MC) simulation, the kinematic features and event selection strategies of nTGCs at FCC-he are investigated by $e^-p \to \nu_{e} \gamma j$, $e^-p \to e^- \gamma j$, $e^-p \to \ell^- \nu_{\ell}\bar{\nu}_{\ell}j$, $e^-p \to \nu_{\ell}\ell^-\ell^+j$, $e^-p \to e^-jjj$ and $e^-p \to \nu_e jjj$ processes, respectively. The expected constraints on the coefficients of the dimension-8 operators as well as the combined constraints are obtained for each process.
	
The reminder of the paper is organized as follows: In Sec.~\ref{level2}, the operators contributing to nTGCs are briefly reviewed.  
We show our numerical results based on MC simulations in Sec.~\ref{level3}. Finally, Sec.~\ref{level4} is a summary.

\section{\label{level2}Dimension-8 operators contributing to nTGCs}
The general dimension-8 SMEFT Lagrangian can be expressed in the following form~\cite{Degrande:2013kka,Gounaris:2000dn,Gounaris:1999kf,Ellis:2019zex}
\begin{equation}
		\begin{split}
		\Delta \mathcal{L}_{{\rm dim-}8}
		&=\sum_{j=1}^{4}\frac{c_j}{\tilde{\Lambda}^4}\mathcal{O}_{j}=\sum_{j=1}^{4}\frac{{\rm sign}(c_j)}{\Lambda^4_j}\mathcal{O}_{j},\\
		\end{split}
		\label{eq.1}
\end{equation}
where the dimensionless coefficients $c_j$ are expected to be around $\mathcal{O}(1)$ and the signs may be ${\rm sign}(c_j)=\pm$. The corresponding ultraviolet cutoff scales are $\Lambda_j\equiv \tilde{\Lambda}/ |c_j|^{1/4}$. The four dimension-8 CP-even effective operators $\mathcal{O}_{j}$ contributing to nTGCs can be expressed as~\cite{Degrande:2013kka}
\begin{equation}
		\begin{split}
			&\mathcal{O}_{\tilde{B}W}=i H^{\dagger}\tilde{B}_{\mu \nu} W^{\mu \rho} \left\{D_{\rho },D^{\nu }\right\}H+h.c.,\;\;\\
			&\mathcal{O}_{B\tilde{W}}=i H^{\dagger} B_{\mu \nu} \tilde{W}^{\mu \rho} \left\{D_{\rho },D^{\nu }\right\}H+h.c.,\\
			&\mathcal{O}_{\tilde{W}W}=i H^{\dagger}\tilde{W}_{\mu \nu} W^{\mu \rho} \left\{D_{\rho },D^{\nu }\right\}H+h.c.,\;\;\\
			&\mathcal{O}_{\tilde{B}B}=i H^{\dagger}\tilde{B}_{\mu \nu} B^{\mu \rho} \left\{D_{\rho },D^{\nu }\right\}H+h.c.,\\
		\end{split}
		\label{eq.2}
\end{equation}
where $H$ denotes the Higgs doublet of the SM and we define the dual field strengths $\tilde{B}_{\mu \nu}= \epsilon_{\mu\nu\alpha\beta} B^{\alpha\beta}$ and $\tilde{W}_{\mu \nu}= \epsilon_{\mu\nu\alpha\beta} W^{\alpha\beta}$, where $W_{\mu\nu}= W_{\mu\nu}^a \sigma^{a}/2$ and $\sigma^a$ are Pauli matrices. For simplicity, we define Wilson coefficients $f_X \equiv \rm{sign}(c_j) / \Lambda^4_j$. 

It has been pointed out that, the four operators in Eqs.~(\ref{eq.1}) and (\ref{eq.2}) do not describe all possible nTGC form factors~\cite{Cepedello:2024ogz,Ellis:2024omd,Murphy:2020rsh,Corbett:2024yoy}, and there are improved operator sets proposed in previous studies~\cite{Ellis:2023ucy,Liu:2024tcz,Ellis:2024omd}.
However, the primary objective of this work is to investigate the sensitivity to nTGCs at $e^-p$ colliders, specifically evaluating the suitability of $e^-p$ colliders for detecting nTGCs in complementary to proton-proton and lepton colliders. 
To ensure comparability with established investigations, we adopt a conventional and widely used operator set that provides a reference benchmark.

\section{\label{level3}Probing nTGCs at \texorpdfstring{$e^{-}p$}{e-p} colliders }
The $e^{-}p$ colliders have cleaner final state and are suitable for discovering new physical signals, accurately measuring some fundamental physical quantities as well as anomalous or flavor-changing couplings. 
To study the effect of nTGCs, a Monte Carlo~(MC) simulation at $5.29$ TeV FCC-he with $\mathcal{L} = 2\;\rm{ab}^{-1}$~\cite{Spor:2021dhf,Gutierrez-Rodriguez:2020gsi} are carried out by using the \verb"MadGraph5_aMC@NLO" toolkit~\cite{Alwall:2014hca,Christensen:2008py}. The \verb"PYTHIA8"~\cite{Sjostrand:2014zea} program is used for parton showering and hadronization and the \verb"DELPHES"~\cite{deFavereau:2013fsa} provides a fast simulation of the LHeC and FCC-he detectors.
The kinematic features are studied using \verb"MLAnalysis"~\cite{Guo:2023nfu}.

\begin{table}[!htbp]
\begin{center}
\begin{tabular}{c|c|c|c|c|c|c}
\hline
channels & $WW\gamma$ & $WWZ$ &  $Z\gamma\gamma$ & $ZZ\gamma$ & $ZZZ$ & $HWW$\\
\hline
$\nu_e \gamma j$ & $\checkmark$ & & & & & \\
\hline
$e^- \gamma j$ &  & & $\checkmark$ & $\checkmark$ & & \\
\hline
$\ell^-\nu_{\ell}\bar{\nu}_{\ell} j$ & $\checkmark$ & $\checkmark$ & $\checkmark$ & $\checkmark$ & $\checkmark$ & $\checkmark$ \\
\hline
$\nu_{\ell}\ell^+\ell^- j$ & $\checkmark$ & $\checkmark$ &  &  &  & $\checkmark$ \\
\hline
$e^-jjj$ & $\checkmark$ & $\checkmark$ &  $\checkmark$ &  $\checkmark$ & $\checkmark$ &  \\
\hline
$\nu _e jjj$ & $\checkmark$ & $\checkmark$ &  &  &  & $\checkmark$ \\
\hline
\hline
operators &  &  &  & & & \\
\hline
$\mathcal{O}_{\tilde{B}W}$ & $\checkmark$ & $\checkmark$ & $\checkmark$ & $\checkmark$ & $\checkmark$ & \\
\hline
$\mathcal{O}_{B\tilde{W}}$ & $\checkmark$ & $\checkmark$ & $\checkmark$ & $\checkmark$ & $\checkmark$ & \\
\hline
$\mathcal{O}_{\tilde{W}W}$ & $\checkmark$ & $\checkmark$ & $\checkmark$ & $\checkmark$ & $\checkmark$ & $\checkmark$ \\
\hline
$\mathcal{O}_{\tilde{B}B}$ & & & $\checkmark$ & $\checkmark$ & $\checkmark$ & \\
\hline
\end{tabular}
\end{center}
\caption{\label{Tab:channels}The correspondence between vertices, processes, and operators.}
\end{table}
The presence of nTGCs can affect the process $e^-p \to \nu_{e} \gamma j$ at tree level.
This process is expected to be sensitive to nTGCs because the neutrino in the final state can avoid the lost of signal events due to the detection efficient of the forward moving leptons, and the photon in the final state introduces no extra electric-weak vertex.
Therefore, the process $e^-p \to \nu_{e} \gamma j$ is of interest in this work.
Since under the influence of nTGCs, the process $e^-p \to \nu_{e} \gamma j$ has only contribution from $WW\gamma$ vertex.
As a complement, the processes $e^-p \to e^- \gamma j$, $e^-p \to \ell^- \nu_{\ell}\bar{\nu_\ell}j$, $e^-p \to {\nu}_{\ell}\ell^-\ell^+j$, $e^-p \to e^-jjj$ and $e^-p \to \nu_e jjj$ are all considered.
Apart from that, it is known that a combined search with multiple channels can enhance the signal of NP, therefore the combined constraints are also investigated. 
A correspondence between the vertices, processes, and operators are shown in Table~\ref{Tab:channels}.

It is well known that the polarization of the primary state electrons in $e^{-}p$ colliders can affect the size of the production cross section. $P(e^-)$ denotes the polarization of electron, ranging from $-80\%$ to $80\%$~\cite{Kuze:2018dqd}. Take the $\mathcal{O}_{\tilde{B}W}$ operator as an example~($f_{\tilde{B}W}$ = 1~${\rm TeV}^{-4}$), the cross section induced by nTGCs with different polarization values are compared in Table~\ref{Tab:polarization}. 
We found that the cross section of the signal can be maximized when the beam polarization is $P(e^-) = -80\%$, and this value is used in the following.

Since the new physics has not yet been observed, we study the phenomenological of the nTGCs in a traditional way, i.e. to study the projected sensitivities.
That is, we consider one operator at a time.
The goal is to investigate whether a process is sensitive to an nTGC operator.
In this way, other operators (dimension-6 or dimension-8 operators contributing to the same processes) are not included.
 


\begin{table}[!htbp]
\begin{center}
\begin{tabular}{c|c|c|c|c|c|c|c|c}
\hline
$P(e^-)$ &$-80\%$ &$-60\%$&$-40\%$&$-20\%$ &$20\%$ &$40\%$ &$60\%$ &$80\%$ \\
\hline
$e^-p \to \nu_{e} \gamma j$	& 24.97 	&  22.21&19.42&16.65	& 11.12 & 8.33 & 5.55 & 2.78    \\
\hline
$e^-p \to e^- \gamma j$	&45.75	&44.83	&43.76	&42.46&40.17	&39.05	&37.96	&36.81\\
\hline
$e^-p \to \ell^- \nu_{\ell}\bar{\nu_\ell}j$	& 2.34	&2.12	&1.83	&1.58 &1.01	&0.83	&0.59	&0.35\\
\hline
$e^-p \to {\nu}_{\ell}\ell^-\ell^+j$	&1.80	&1.66	&1.51	&1.35 &1.08	&0.93	&0.80	&0.71	\\
\hline
$e^-p \to e^-jjj$	&271.19	&264.14	&257.49	&251.55 &236.55	&230.23	&222.03	&214.76\\
\hline
$e^-p \to \nu_e jjj$	&108.88	&96.66	&84.23	&72.56	&48.29	&36.34	&23.95	&12.05\\
\hline
\end{tabular}
\end{center}
\caption{\label{Tab:polarization}Take the $\mathcal{O}_{\tilde{B}W}$ operator as an example, the values of cross section (fb) contributed by nTGCs under different polarizations at $5.29$ TeV FCC-he with $\mathcal{L} = 2\;\rm{ab}^{-1}$.}
\end{table}

\subsection{\label{sec3.1}The process \texorpdfstring{$e^-p \to \nu _{e} \gamma j$}{e-p to v a j}}

\begin{figure}[!htbp]
\centering
\includegraphics[width=0.45\hsize]{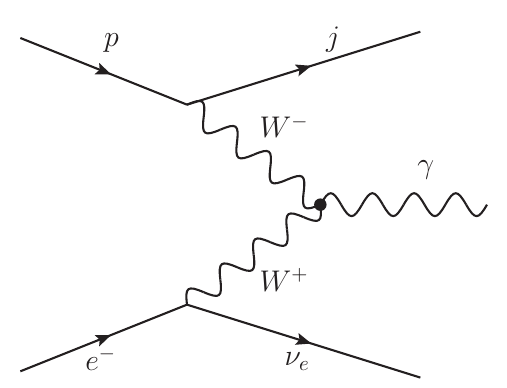}\\
\includegraphics[width=0.45\hsize]{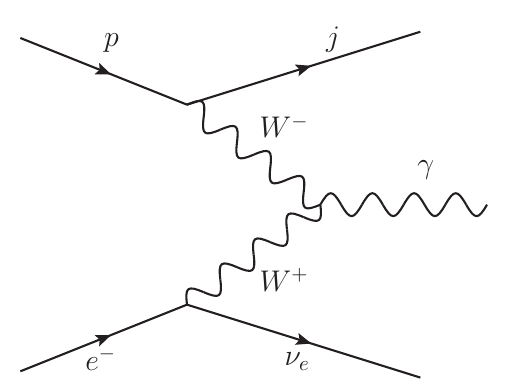}
\includegraphics[width=0.45\hsize]{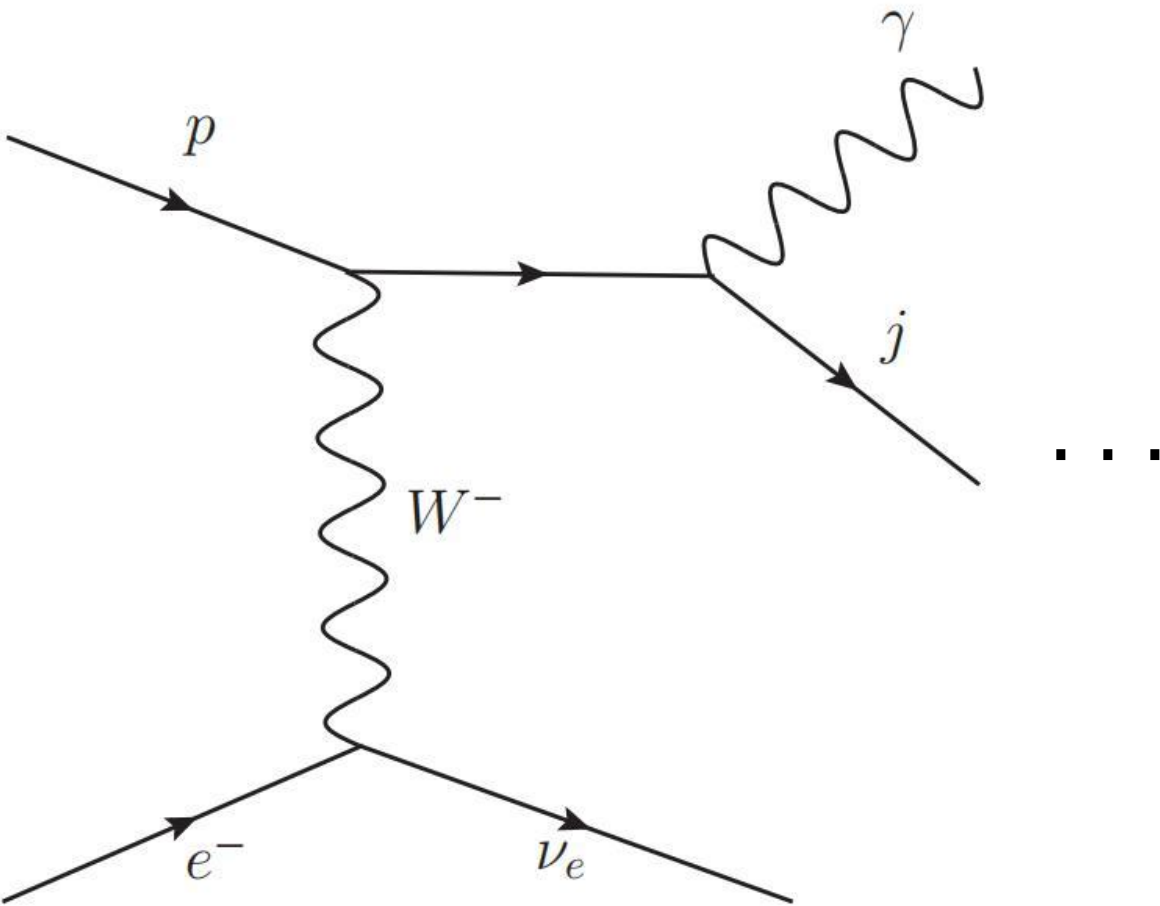}
\caption{\label{fig:epvja}Typical Feynman diagrams of nTGCs contribution~(the first row) and the SM backgrounds~(the second row) for the process $e^-p\to \nu _e \gamma j$.}
\end{figure}

The $\nu _{e} \gamma j$ channel can be contributed by the $WW \gamma$ vertex. 
At tree level, the typical Feynman diagrams are shown in Fig.~\ref{fig:epvja}.
For the $e^-p \to \nu _{e} \gamma j$ process, we require the particle numbers in the final states to be $N_{jet}\geq 1$ and $N_{\gamma}\geq 1$, where $N_{jet}$, and $N_{\gamma}$ are the number of outgoing jets and photons, respectively.

\begin{figure}[!htbp]
	\centering{
		\includegraphics[width=0.45\hsize]{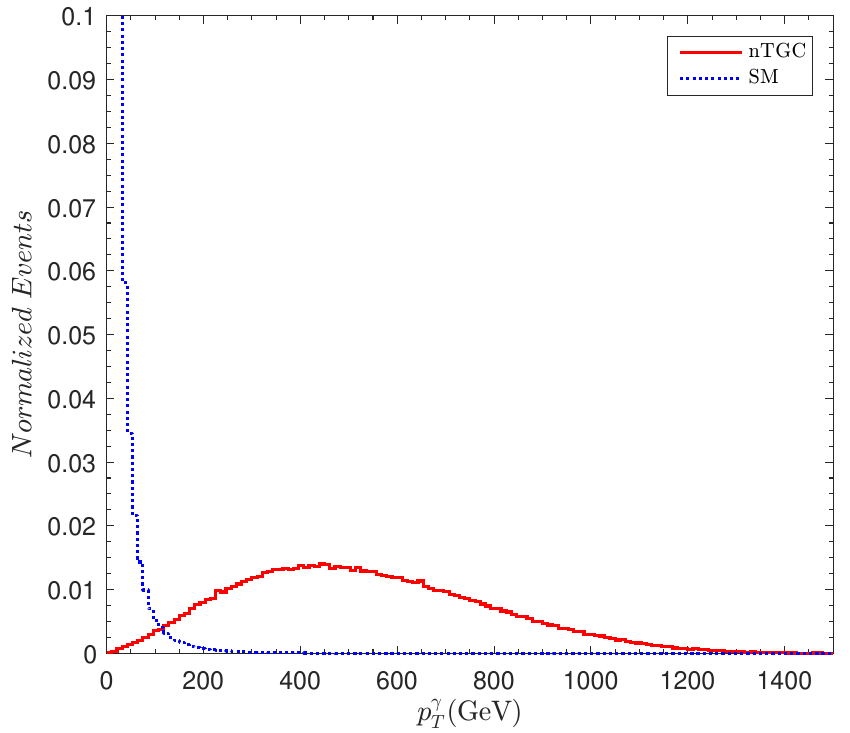}
		\includegraphics[width=0.45\hsize]{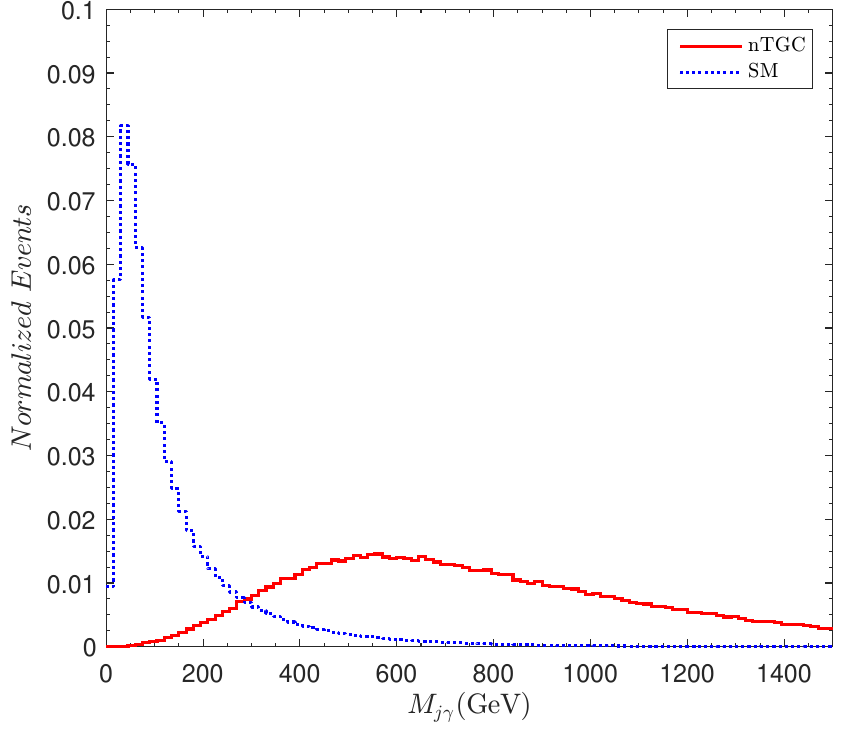}\\
		\includegraphics[width=0.45\hsize]{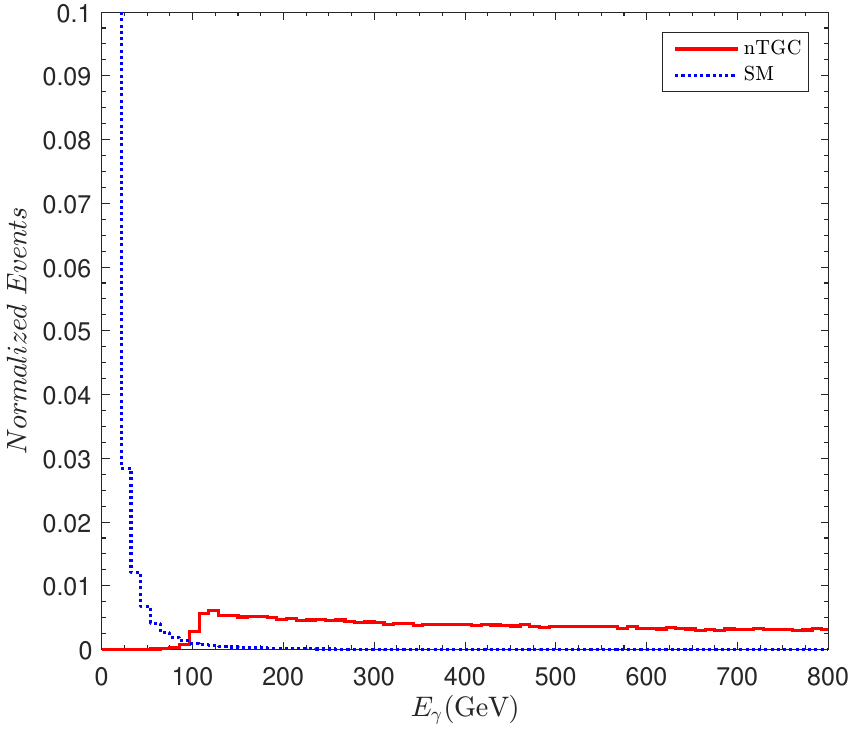}
		\caption{\label{Fig:cut3.1}The normalized distributions of $p_T^{\gamma}$, $M_{j\gamma}$ and $E_{\gamma}$ for $e^-p \to \nu _{e} \gamma j$.}}
\end{figure}
It can be seen that, for the SM background, the photon is typically radiated from one of the beams, which is different from the case of nTGCs.
As a consequence, it can be expected that, for the background, the photon is collinear to the beams, which results in a small transverse momentum of hardest photon~(denoted as $p_T^{\gamma}$), and a relatively small $M_{j\gamma}$~(the invariant mass of the hardest photon and the hardest jet). 
The normalized distributions of these kinematic variables at $5.29$ TeV FCC-he with $\mathcal{L} = 2\;\rm{ab}^{-1}$ are shown in Fig.~\ref{Fig:cut3.1}.
As shown in Fig.~\ref{Fig:cut3.1}, the $E_{\gamma}$, $p_T^{\gamma}$ and $M_{j\gamma}$ distributions for the
SM background are concentrated in the low mass range. In contrast, the signals are primarily concentrated in higher mass range.
Depending on the characteristics of these kinematic distributions, different optimized kinematic cuts are applied to reduce background and increase statistical significance, in particular $p_T^{\gamma}\geq 250\;\rm{GeV}$, $M_{j\gamma} \geq 700\;\rm{GeV}$ and $E_{\gamma} \geq 350\;\rm{GeV}$.

\begin{table}[!htbp]
	\begin{center}
		\begin{tabular}{c|c|c}
			\hline
			& Signal (pb) & Background (pb)  \\
			\hline
			 $N_{jet}\geq 1$, $N_{\gamma}\geq 1$	& 0.125  	&  0.336   \\
			\hline
			$p_T^{\gamma}\geq 250\;\rm{GeV}$	& 0.117  	&0.0032  \\
			\hline
			$M_{j\gamma} \geq 700\;\rm{GeV}$	& 0.099   	& 0.0019  \\
			\hline
			$E_{\gamma} \geq 350\;\rm{GeV}$	&   0.096 	&0.00015   \\
			\hline
		\end{tabular}
	\end{center}
	\caption{\label{Tab:cross3.1}The production cross sections of the signal and SM background after the improved
		cuts at $5.29$ TeV FCC-he with $\mathcal{L} = 2\;\rm{ab}^{-1}$. }
\end{table}
Take the $\mathcal{O}_{\tilde{B}W}$ operator as an example, the production cross sections of the signal and background after imposing improved cuts at $5.29$ TeV FCC-he with $\mathcal{L} = 2\;\rm{ab}^{-1}$ are given in Table~\ref{Tab:cross3.1}.
The results are calculated with $f_{\tilde{B}W}$ = 1~${\rm TeV}^{-4}$.
As can be seen that the background is strongly suppressed, while the signal still has good efficiency after cuts.

When the effect of the interference term is considered, the total cross sections after event selection strategy~(denoted by $\sigma_{\rm{nTGC}}$) can be fitted as a bilinear function of the coefficients,
\begin{equation}
	\begin{split}
		&\sigma_{\rm{nTGC}} =\sigma_{\rm{SM}} + \sigma_{\rm{INT}} + \sigma_{\rm{NP}}.
	\end{split}
	\label{eq.3.1.1}
\end{equation}
 
$\sigma_{\rm{NP}} =  f^2 \times \widehat{\sigma}_{\rm{np}}$, where $\widehat{\sigma}_{\rm{np}}$ is NP contribution after cuts to be fitted and $f$ is the coefficient.
$\sigma_{\rm{INT}} = f \times \widehat{\sigma}_{\rm{int}}$, where $\widehat{\sigma}_{\rm{int}}$	 is the interference parameter to be fitted. 
The interference is taken into account when studying the expected constraints.

\begin{figure}[!htbp]
	\centering{
		\includegraphics[width=0.6\hsize]{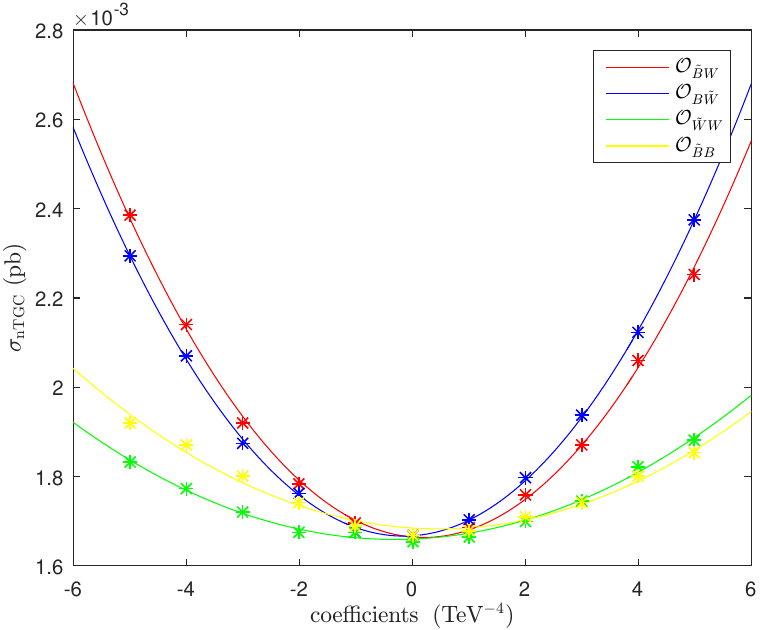}
		\caption{\label{vja1234}The total cross sections after cuts as functions of operator coefficients for the process $e^-p\to \nu_e\gamma j$.}}
\end{figure}
The total cross sections after cuts are obtained by scanning the values of coefficients $f_{\tilde{B}W}$, $f_{B\tilde{W}}$, $f_{\tilde{W}W}$ and $f_{\tilde{B}B}$ for one operator at a time.
The results of the fittings according to Eq.~(\ref{eq.3.1.1}) are shown in Fig.~\ref{vja1234}. 
It can be seen that the bilinear function fits well.

The expected constraints on nTGCs coefficients at FCC-he can be estimated by using signal significance defined as
\begin{equation}
	\begin{split}
		\mathcal{S}_{stat} =\frac{N_{\rm{nTGC}}}{\sqrt{N_{\rm{nTGC}}+N_{\rm{SM}}}}
	\end{split}
	\label{eq.3.1.2}
\end{equation}
where $N_{\rm{nTGC}}=\mathcal{L}(\sigma_{\rm{INT}} + \sigma_{\rm{NP}})$, $N_{\rm{SM}}=\mathcal{L}\sigma_{\rm{SM}}$. 
$\mathcal{S}_{stat}$ is a function of $f$, and then the expected constraints at $\mathcal{S}_{stat}=2,3$ and $5$ can be obtained by solving the equations accordingly.
The results are resented in Table~\ref{Tab:constraints} at the end of this section.

In addition, to enhance the reliability of our conclusions, taking $e^-p \to \nu _{e} \gamma j$ process as an example, we also investigated the sensitivity of the coefficient $f_{\tilde{B}W}$ at $1.30$ TeV LHeC with $\mathcal{L} = 1\;\rm{ab}^{-1}$, which are $[-62.29\;{\rm TeV}^{-4}, 21.75\;{\rm TeV}^{-4}]$, $[-71.28\;{\rm TeV}^{-4}, 30.74\;{\rm TeV}^{-4}]$ and $[-88.37\;{\rm TeV}^{-4}, 47.82\;{\rm TeV}^{-4}]$, corresponding to $\mathcal{S}_{stat}=2,3$ and $5$, respectively. For $3.50$ TeV FCC-he with $\mathcal{L} = 2\;\rm{ab}^{-1}$, the expected constraints for $f_{\tilde{B}W}$ are $[-4.34\;{\rm TeV}^{-4}, 5.37\;{\rm TeV}^{-4}]$, $[-5.44\;{\rm TeV}^{-4}, 6.47\;{\rm TeV}^{-4}]$ and $[-7.18\;{\rm TeV}^{-4}, 8.22\;{\rm TeV}^{-4}]$, respectively.

\subsection{\label{sec3.2}The process \texorpdfstring{$e^-p \to e^- \gamma j$}{e-p to e- a j}}

\begin{figure}[!htbp]
\centering
\includegraphics[width=0.45\hsize]{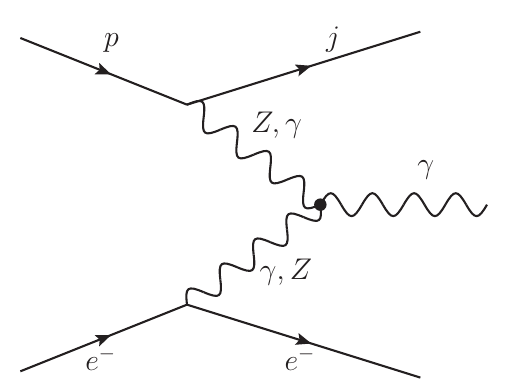}\\
\includegraphics[width=0.45\hsize]{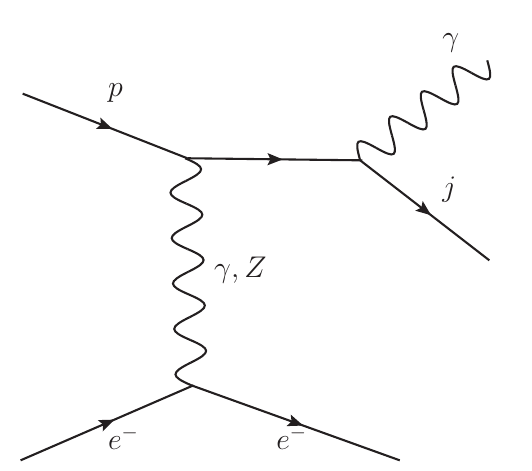}
\includegraphics[width=0.45\hsize]{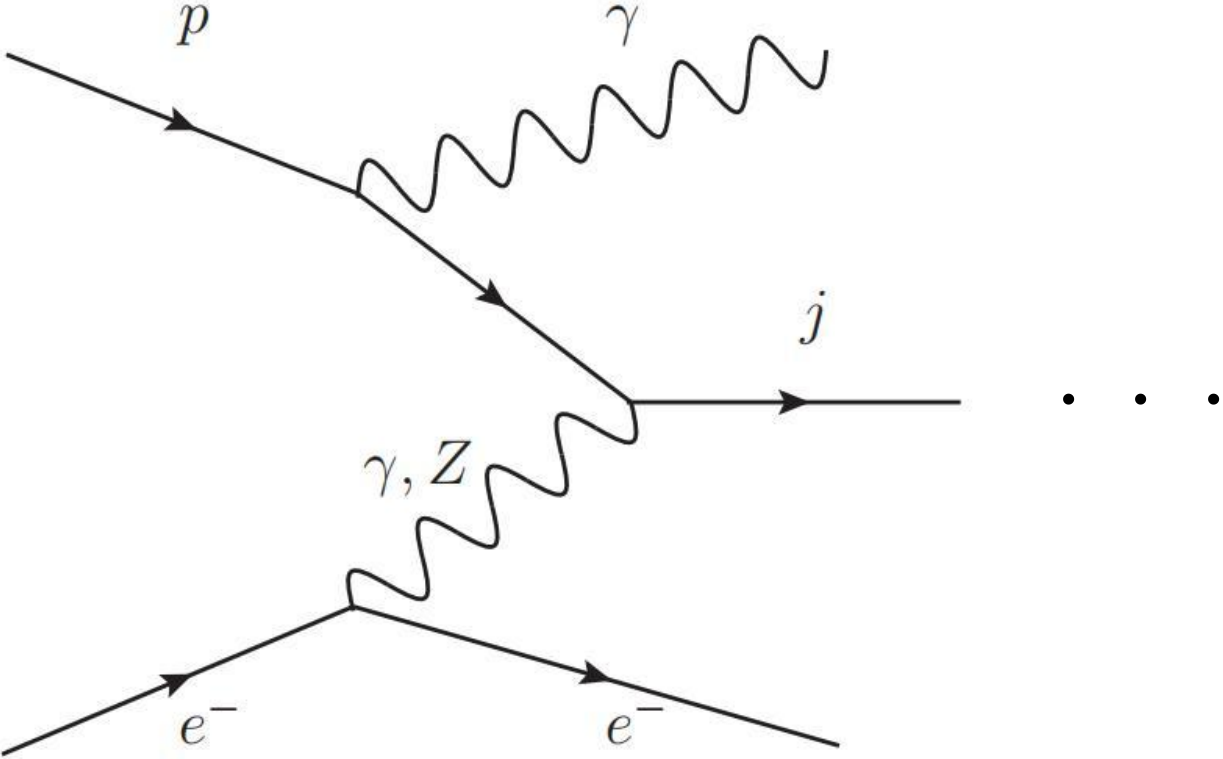}
\caption{\label{fig:epeja}Typical Feynman diagrams of nTGCs contribution~(the first row) and the SM backgrounds~(the second row) for the process $e^-p \to e^- \gamma j$.}
\end{figure}
The $e^- \gamma j$ channel can be contributed by the $ZZ\gamma$ and $\gamma\gamma\gamma$ vertices.
The typical Feynman diagrams are shown in Fig.~\ref{fig:epeja}.
For the characterization, we require the particle numbers in the final states to be $N_{jet}\geq 1$ and $N_{\gamma}\geq 1$.

\begin{figure}[!htbp]
	\centering{
		\includegraphics[width=0.45\hsize]{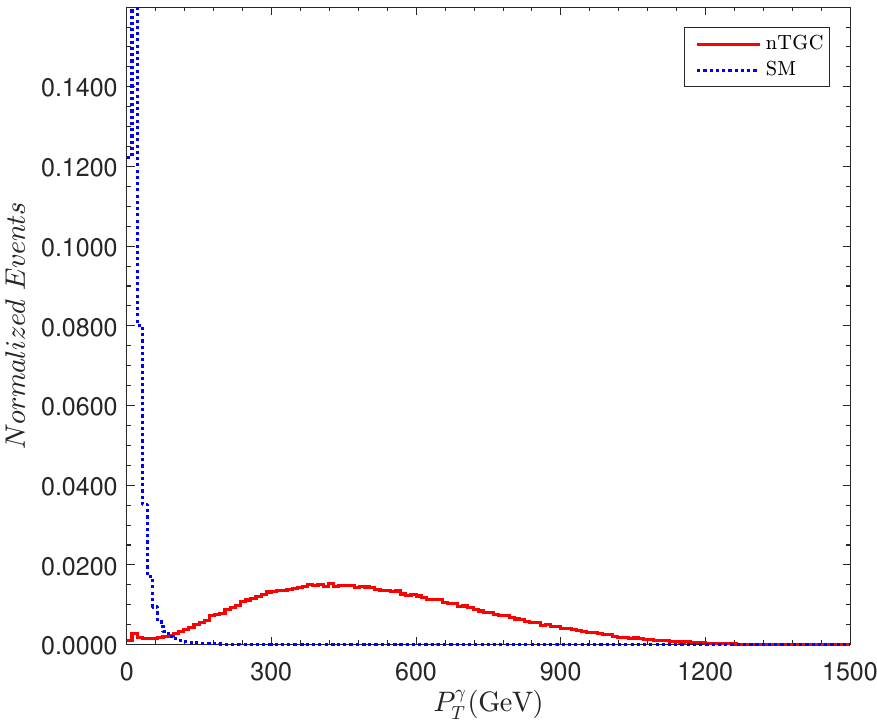}
		\includegraphics[width=0.45\hsize]{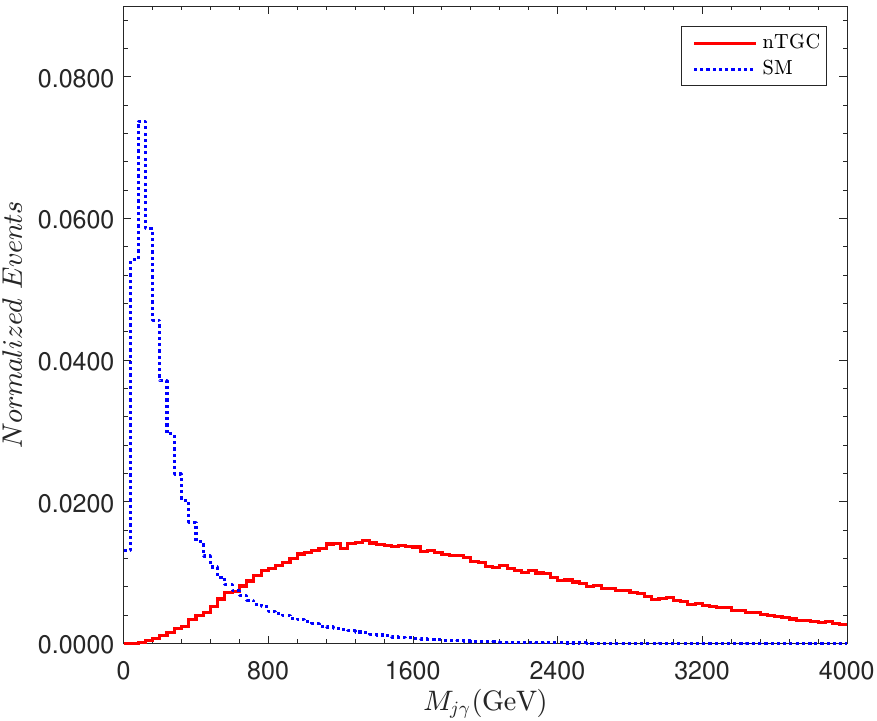}\\
		\includegraphics[width=0.45\hsize]{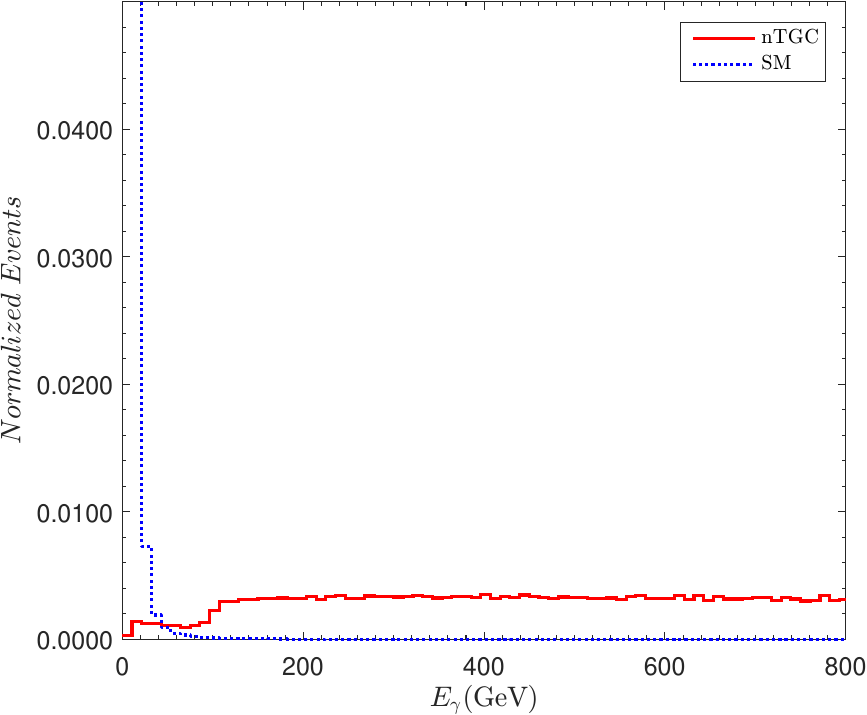}
		\includegraphics[width=0.45\hsize]{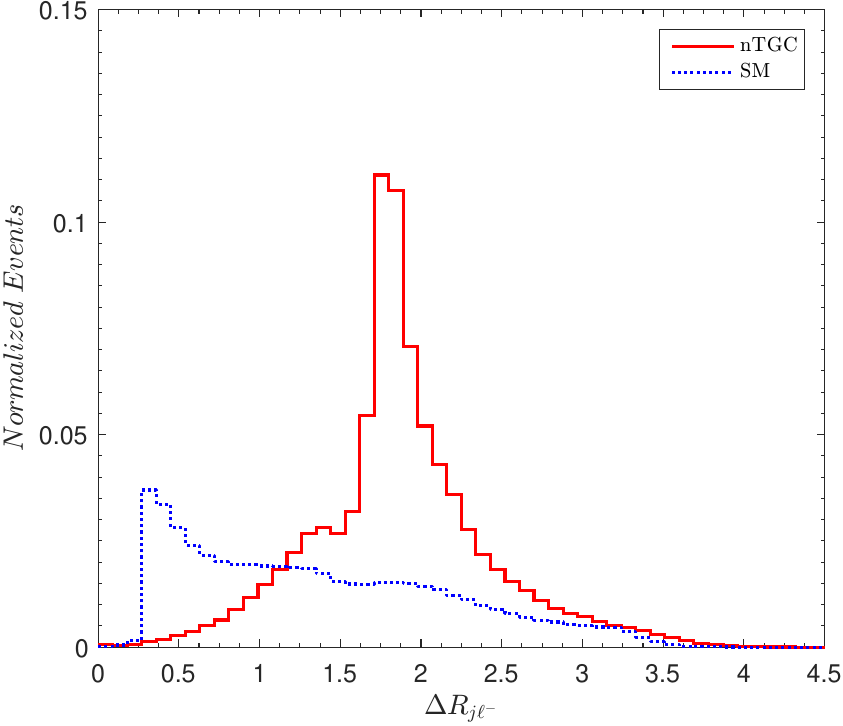}\\
		\includegraphics[width=0.45\hsize]{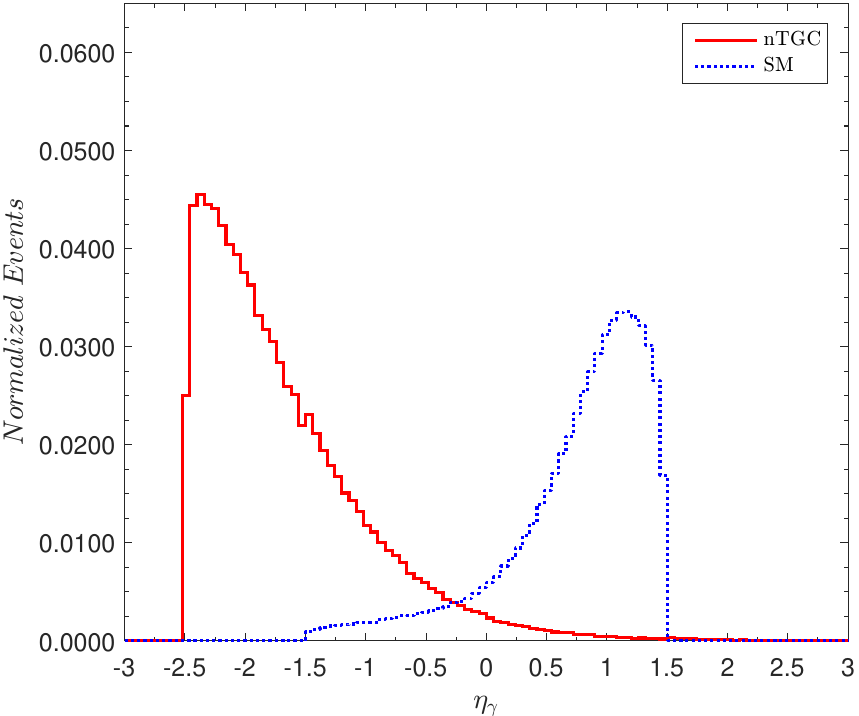}
		\caption{\label{Fig:cut3.2}The normalized distributions of $p_T^{\gamma}$, $M_{j\gamma}$, $E_{\gamma}$, $\Delta R_{ j\ell^-}$ and $\eta_{\gamma}$  for $e^-p \to e^- \gamma j$.}}
\end{figure}

It can be seen that, the Feynman diagrams are similar to the ones for the process $e^-p \to \nu \gamma j$.
As a consequence, $p_T^{\gamma}$, $M_{j\gamma}$ and $E_{\gamma}$ can also been used for event selection. We present in Fig.~\ref{Fig:cut3.2} the distributions of $p_T^{\gamma}$, $M_{j\gamma}$, $E_{\gamma}$ and $\Delta R_{ j\ell^-}$ at $5.29$ TeV FCC-he with $\mathcal{L} = 2\;\rm{ab}^{-1}$.
It can be expected that $M_{e^-\gamma}$ should also be large for the signal events.
However, a large part of the electrons move along the z-axis which are difficult to be detected, as a result $M_{e^- \gamma}$ is not considered. 
The angular separation requirements are $\Delta R_{ j\ell}$ for jets and leptons, with $\Delta R$ defined as $\sqrt{(\Delta \phi)^2+(\Delta \eta)^2}$, where $\Delta \phi$ and $\Delta \eta$ are the azimuth difference and pseudorapidity between a jet and a lepton.
Additionally, the pseudovelocity of photons $\eta_{\gamma}$ is also chosen to enhance signal efficiency. As shown in Fig.~\ref{Fig:cut3.2},  due to the relatively high momentum of the final-state particles, the leptons and jets from the $e^- \gamma j$ decay are typically very distant, resulting in large angular separations in signal events. In contrast, background events often have narrower angular distributions.
The kinematic cuts are chosen as $p_T^{\gamma}\geq 200\;\rm{GeV} $, $M_{j\gamma} \geq 600\;\rm{GeV}$, $E_{\gamma} \geq 350\;\rm{GeV}$, $\Delta R_{ j\ell^-}\geq 1.2$ and $-3\leq \eta_{\gamma} \leq -1$, respectively.

\begin{table}[!htbp]
	\begin{center}
		\begin{tabular}{c|c|c}
			\hline
			 & Signal (pb) & Background (pb)  \\
			\hline
		 $N_{jet}\geq 1$, $N_{\gamma}\geq 1$	& 0.127  	&  0.257   \\
			\hline
			$p_T^{\gamma}\geq 200\;\rm{GeV} $	& 0.125  	&0.0011  \\
			\hline
			$M_{j\gamma} \geq 600\;\rm{GeV}$	& 0.105   	& 0.00073  \\
			\hline
			$E_{\gamma} \geq 350\;\rm{GeV}$ 	&   0.103 	&0.00036   \\
			\hline
		$\Delta R_{ j\ell^-}\geq 1.2$	&  0.102 	&0.00025  \\
			\hline
		$-3 \le \eta_{\gamma} \le -1$	& 	 0.101 	&0.00015  \\
			\hline
		\end{tabular}
	\end{center}
	\caption{\label{Tab:cross3.2}After different cuts applied, the cross sections for the signal and SM background at $5.29$ TeV FCC-he with $\mathcal{L} = 2\;\rm{ab}^{-1}$.}
\end{table}
We show the production cross sections of the signal and background at $5.29$ TeV FCC-he with $\mathcal{L} = 2\;\rm{ab}^{-1}$ after taking the cuts in Table~\ref{Tab:cross3.2}. The results are calculated with $f_{\tilde{B}W}$ = 1~${\rm TeV}^{-4}$.
It can be seen that the background is effectively suppressed.

\begin{figure}[!htbp]
	\centering{
		\includegraphics[width=0.6\hsize]{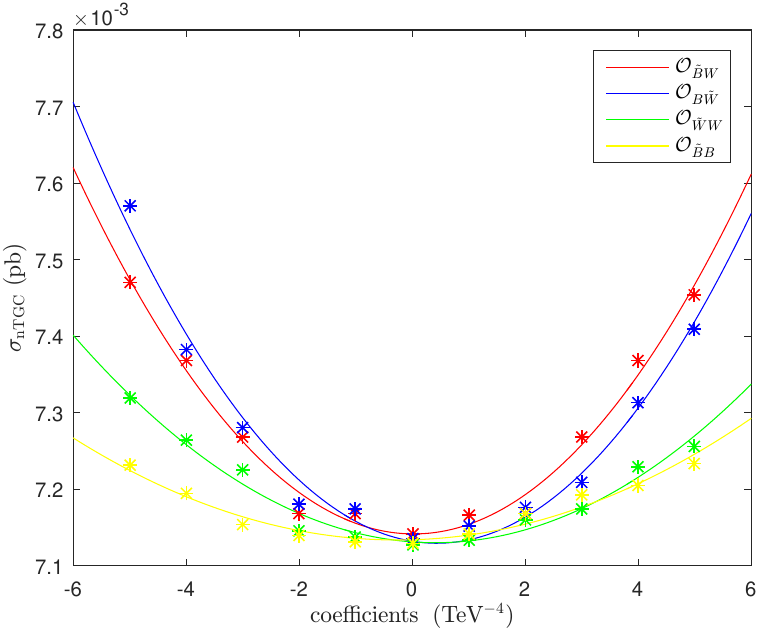}
		\caption{\label{Fig:fit3.2}Same as Fig.~\ref{vja1234} but for $e^-p \to e^- \gamma j$.}}
\end{figure}
The cross sections after cuts are also fitted, as shown in Fig.~\ref{Fig:fit3.2}. 
By using signal significance, the expected constraints for $f_{\tilde{B}W}$ corresponding to $\mathcal{S}_{stat}=2,3$ and $5$ are $[-2.00\;{\rm TeV}^{-4}, 2.64\;{\rm TeV}^{-4}]$, $[-2.85\;{\rm TeV}^{-4}, 2.99\;{\rm TeV}^{-4}]$ and $[-4.12\;{\rm TeV}^{-4}, 4.33\;{\rm TeV}^{-4}]$, respectively.

\subsection{\label{sec3.3}The process \texorpdfstring{$e^-p \to \ell^- \nu_{\ell}\bar{\nu_\ell}j$}{e-p to l- v j}}

\begin{figure}[!htbp]
\centering
\includegraphics[width=0.4\hsize]{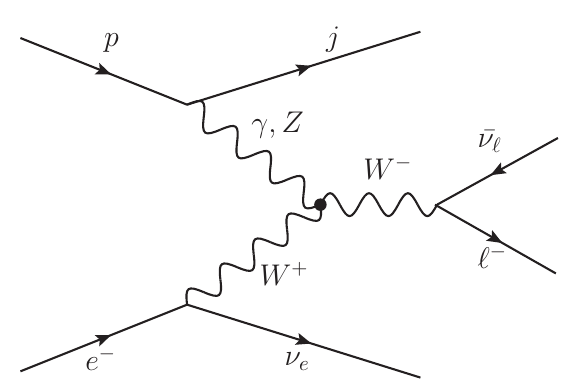}
\includegraphics[width=0.4\hsize]{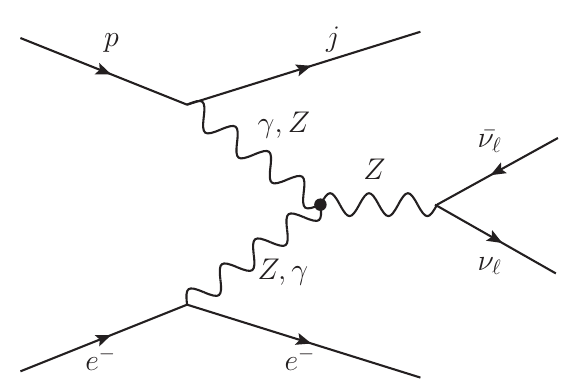}\\
\includegraphics[width=0.43\hsize]{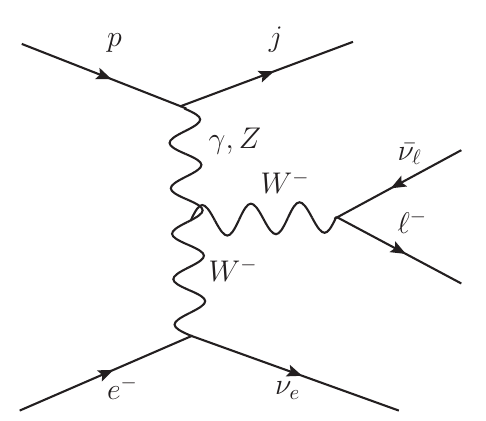}
\includegraphics[width=0.4\hsize]{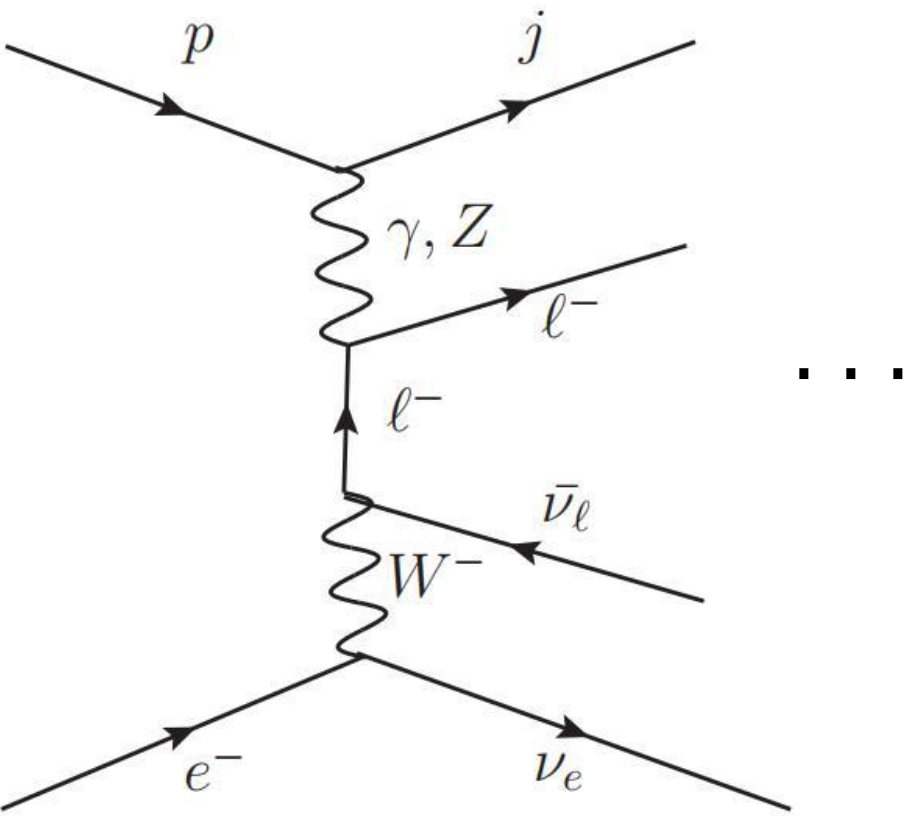}
\caption{\label{fig:eplvvj}Typical Feynman diagrams of nTGCs contribution~(the first row) and the SM backgrounds~(the second row) for the process $e^-p \to \ell^- \nu_{\ell}\bar{\nu_\ell}j$.}
\end{figure}

The typical Feynman diagrams for the process $e^-p \to \ell^- \nu_{\ell}\bar{\nu_\ell}j$ are shown in Fig.~\ref{fig:eplvvj}. 
All numerical results are obtained after applying the particle number cuts, $N_{\ell^-}\geq 1$ and $N_{jet}\geq 1$.

\begin{figure}[!htbp]
	\centering{
		\includegraphics[width=0.45\hsize]{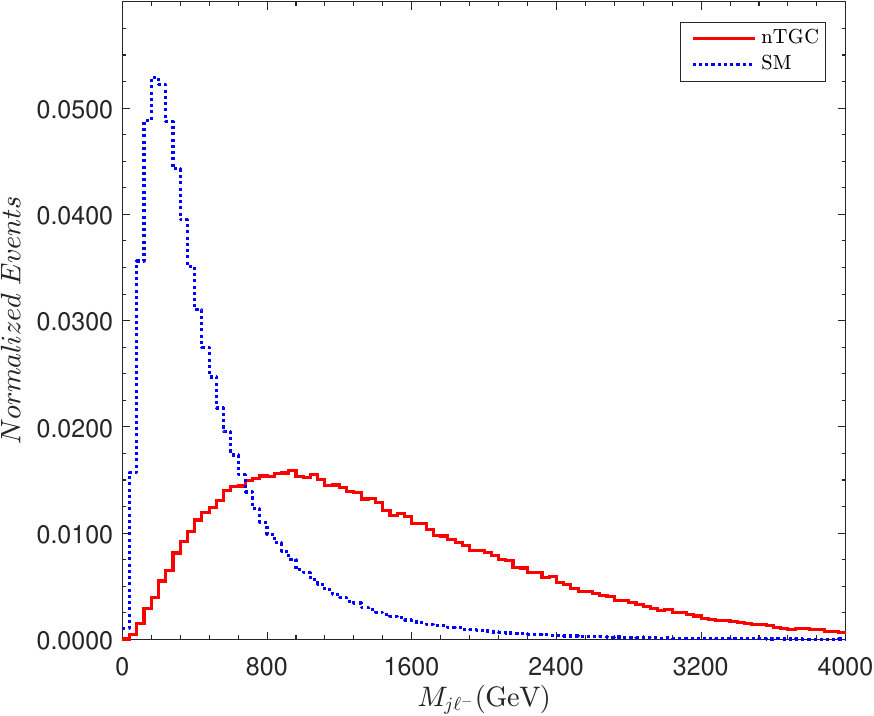}
		\includegraphics[width=0.45\hsize]{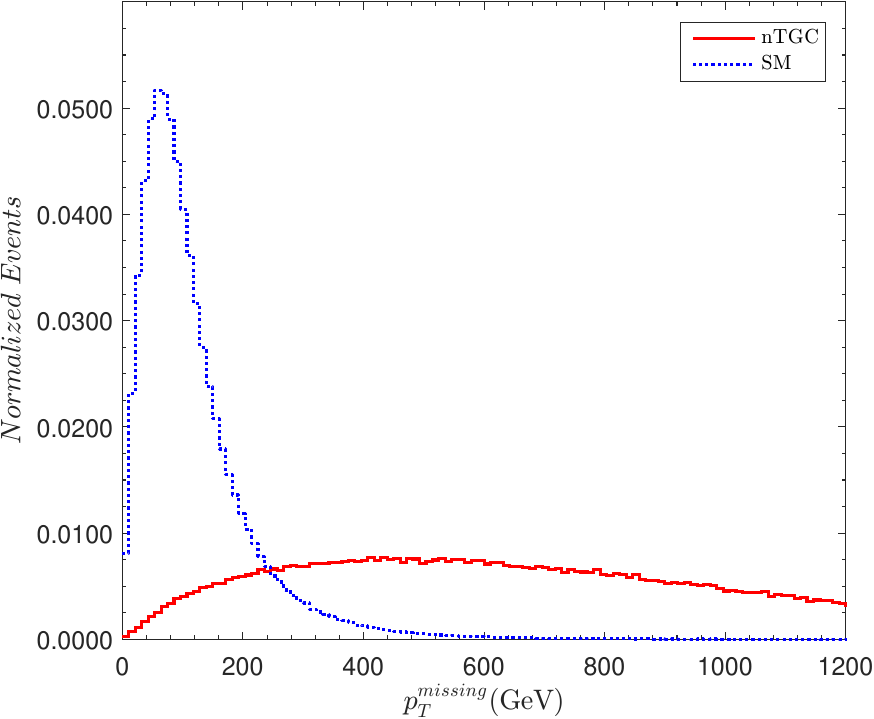}\\
		\includegraphics[width=0.45\hsize]{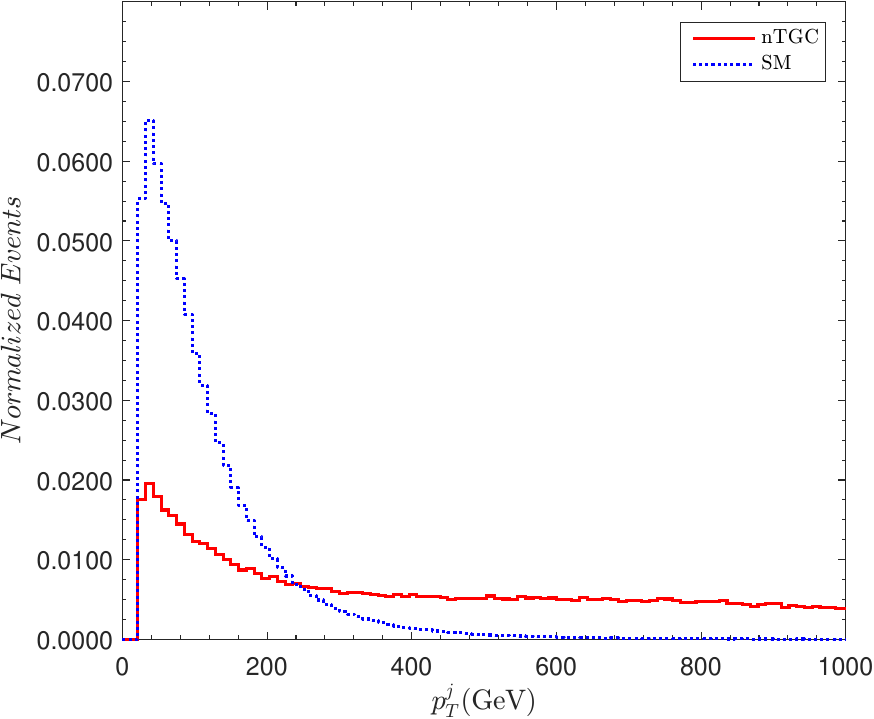}
		\caption{\label{Fig:cut3.3}The normalized distributions of $M_{j\ell^-}$, $p_T^{missing}$ and $p_T^{j}$ for $e^-p \to \ell^- \nu_{\ell}\bar{\nu_\ell}j$.}}
\end{figure}
\begin{table}[!htbp]
	\begin{center}
		\begin{tabular}{c|c|c}
			\hline
			& Signal (pb) & Background (pb)  \\
			\hline
			$N_{\ell^-}\geq 1$, $N_{jet}\geq 1$	& 0.111  	&  0.343   \\
			\hline
		 $M_{j \ell^-} \geq 500\;\rm{GeV}$	& 0.083  	&0.059  \\
			\hline
			$p_T^{\rm{missing}} \geq 300\;\rm{GeV}$	& 0.071   	& 0.0057  \\
			\hline
			$P^{j}_T\geq 350\;\rm{GeV}$ 	&   0.049 	&0.0031   \\
			\hline
		\end{tabular}
	\end{center}
	\caption{\label{Tab:cross3.3}After different cuts, the cross sections for the signal and SM background for the $e^-p \to \ell^- \nu_{\ell}\bar{\nu_\ell}j$ process at $5.29$ TeV FCC-he with $\mathcal{L} = 2\;\rm{ab}^{-1}$.}
\end{table}
As seen in Fig.~\ref{fig:eplvvj}, the $Z$ and $W$ bosons in the signal are energetic, and then the energy of the generated leptons is expected to be more energetic compared to the SM background. 
Therefore, $p_T^{\rm{missing}}$~(the missing transverse momentum) should be large for the signal.
The charged leptons in the final state are either from $W$ bosons or the residue of the beam, so a large $M_{j\ell}$ can be expected.
In addition, the typical transfer momentum for the signal is larger, a tail in the high energy region can be expected for $P^j_T$, which is the transverse momentum of hardest jet.
The normalized distributions of $p_T^{\rm{missing}}$, $M_{j\ell}$ and $P^{j}_T$ are shown in Fig.~\ref{Fig:cut3.3}.
We cut off these events with $M_{j \ell^-} \geq 500\;\rm{GeV}$, $p_T^{\rm{missing}} \geq 300\;\rm{GeV}$ and $P^{j}_T\geq 350\;\rm{GeV}$ in order to suppress the background.
The cross sections after cuts are shown in Table~\ref{Tab:cross3.3}~($f_{\tilde{B}W}$ = 1~(${\rm TeV}^{-4}$)).

\begin{figure}[!htbp]
	\centering{
		\includegraphics[width=0.6\hsize]{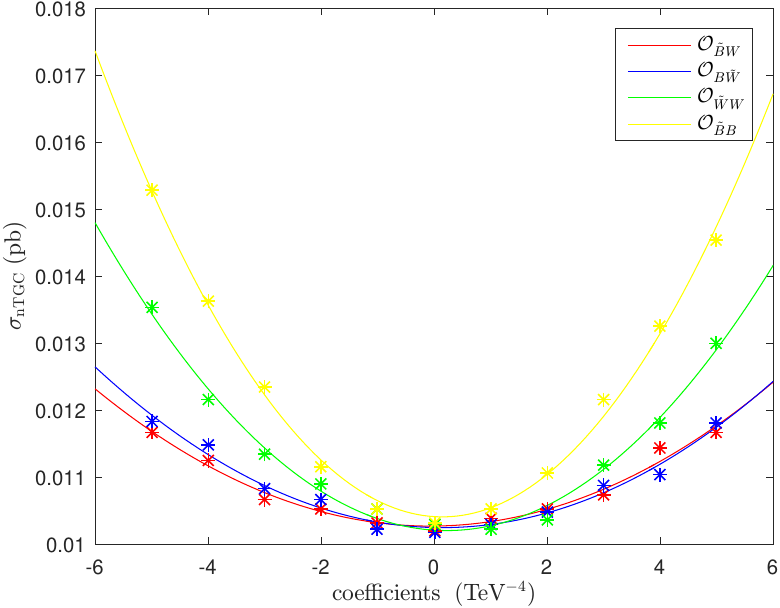}
		\caption{\label{Fig:fit3.3}Same as Fig.~\ref{vja1234} but for $e^-p \to \ell^-\nu_{\ell}\bar{\nu_\ell}j$.}}
\end{figure}
Total cross sections after cuts are fitted and shown in Fig.~\ref{Fig:fit3.3}.
The expected constraints for different signal significance are summarized at the end of this section in Table~\ref{Tab:constraints}.

\subsection{\label{sec3.4}The process \texorpdfstring{$e^-p \to \nu_{\ell}\ell^- \ell^+ j$}{e-p to v l- l+ j}}

\begin{figure}[!htbp]
\centering
\includegraphics[width=0.4\hsize]{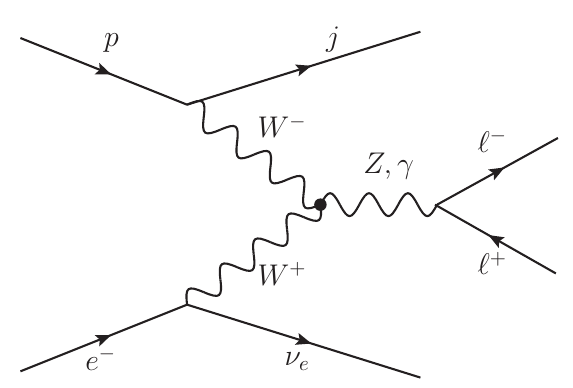}
\includegraphics[width=0.4\hsize]{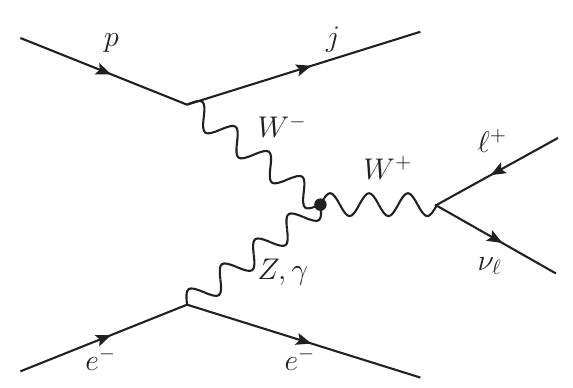}\\
\includegraphics[width=0.43\hsize]{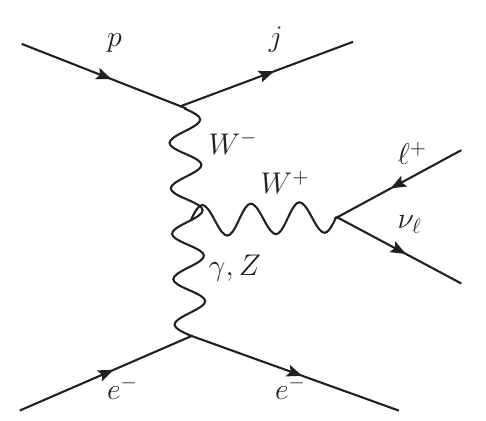}
\includegraphics[width=0.4\hsize]{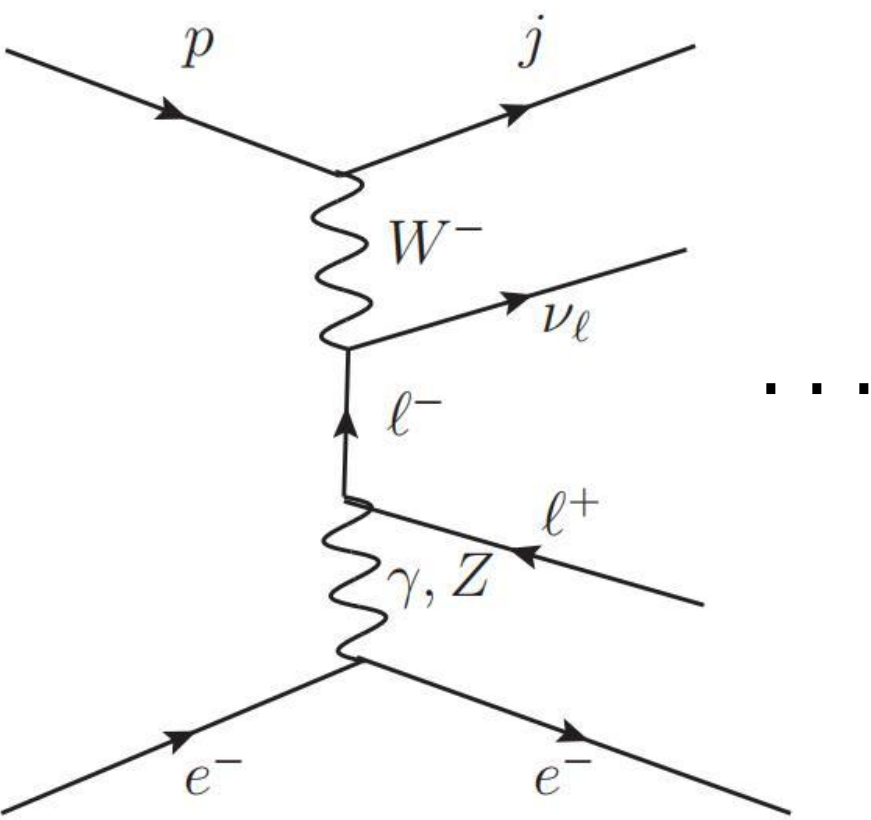}
\caption{\label{fig:epvllj}Typical Feynman diagrams of nTGCs contribution~(the first row) and the SM backgrounds~(the second row) for the process $e^-p\to \nu _{\ell} \ell^+\ell^-j$.}
\end{figure}
The Feynman diagrams of the signals and backgrounds for the process $e^-p \to \nu _{\ell} \ell^- \ell^+ j$ are shown in Fig.~\ref{fig:epvllj}. 
The final-states are required to satisfy $N_{jet}\geq 1$. As shown in Fig.~\ref{fig:epvllj}~(the first row), the positive and negative leptons in the final state tend to be co-linear and cannot be distinguished due to the high energy of the $Z$ boson. In order not to depress the signal, $N_{\ell}\geq 1$ is required. 

The kinematic distributions for the signal and background are similar to the previous subsection, so we apply the $M_{j \ell}$~(the invariant mass of the hardest jet and the hardest outgoing charged lepton), $p^{\ell}_T$~(the transverse momentum of hardest outgoing charged lepton) and $p^{\rm{missing}}_T$~(the missing transverse momentum) in this case.

\begin{figure}[!htbp]
	\centering{
		\includegraphics[width=0.45\hsize]{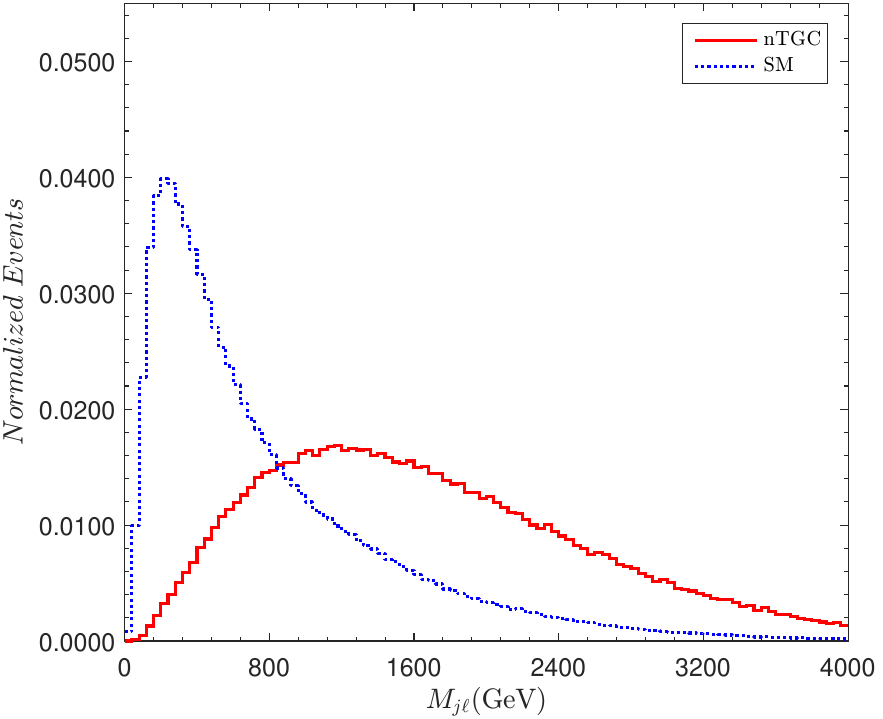}
		\includegraphics[width=0.45\hsize]{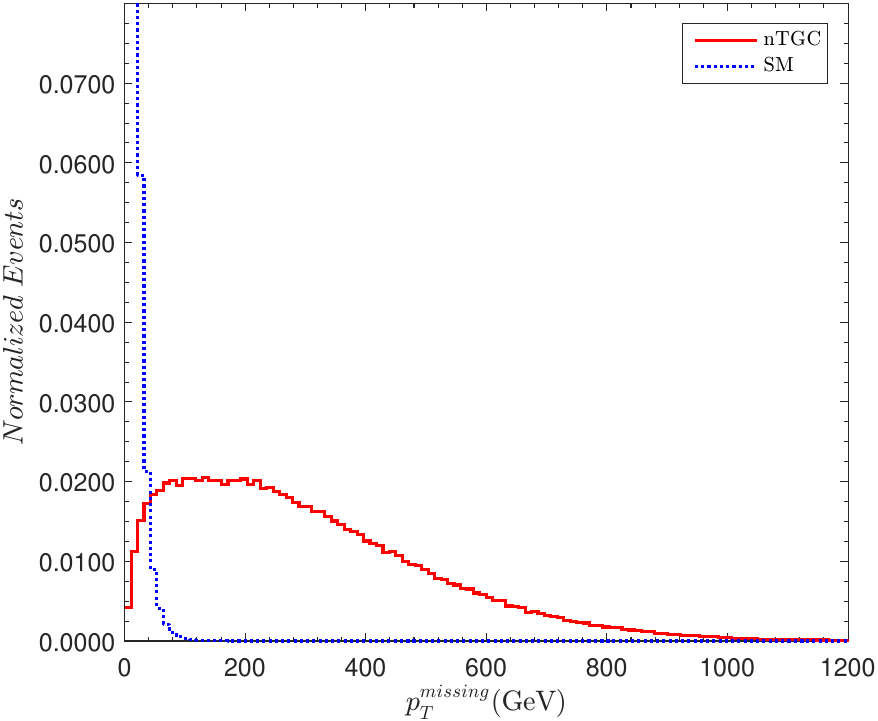}
		\includegraphics[width=0.45\hsize]{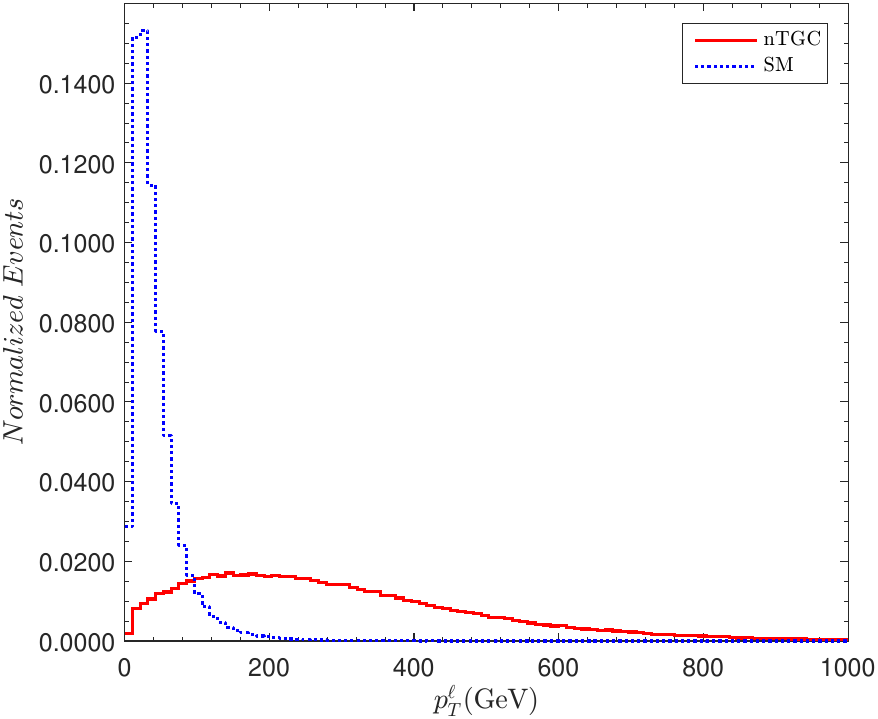}
		\caption{\label{Fig:cut4}The normalized distributions of $M_{j\ell}$, $p_T^{\rm{missing}}$ and $p_T^{\ell}$ for $e^-p \to \nu_{\ell}\ell^- \ell^+ j$.}}
\end{figure}
\begin{table}[!htbp]
	\begin{center}
		\begin{tabular}{c|c|c}
			\hline
			& Signal (pb) & Background (pb)  \\
			\hline
			 $N_{jet}\geq 1$. $N_{\ell}\geq 1$	& 0.136  	&  0.417   \\
			\hline
			$M_{j \ell} \geq 600\;\rm{GeV}$	& 0.077  	&0.028  \\
			\hline
			$p_T^{\rm{missing}} \geq 250\;\rm{GeV}$	& 0.043   	& 0.0015  \\
			\hline
			$P^{\ell}_T\geq 250\;\rm{GeV}$	&   0.033 	&0.0003   \\
			\hline
		\end{tabular}
	\end{center}
	\caption{\label{Tab:cross4}After different cuts, the cross sections for the signal and SM background for the $e^-p \to \nu_{\ell}\ell^- \ell^+ j$ process at $5.29$ TeV FCC-he with $\mathcal{L} = 2\;\rm{ab}^{-1}$.}
\end{table}
The normalized distributions of these kinematic variables are shown in Fig.~\ref{Fig:cut4}.
As shown in Fig.~\ref{Fig:cut4}, for the SM background, the distribution peaks at lower values of these kinematic variables, while for the signal, it shifts to relatively higher values. According to the features of these kinematic distributions, different optimized kinematical cuts are applied to reduce background and improve the statistical significance, which are $M_{j \ell} \geq 600\;\rm{GeV}$, $P^{\ell}_T\geq 250\;\rm{GeV}$ and $p_T^{\rm{missing}} \geq 250\;\rm{GeV}$.
In Table~\ref{Tab:cross4}, we summarized the cross sections of the signal and background after imposing improved cuts and the characteristic parameter $f_{\tilde{B}W}$ = 1~${\rm TeV}^{-4}$. 

\begin{figure}[!htbp]
 	\centering{
 		\includegraphics[width=0.6\hsize]{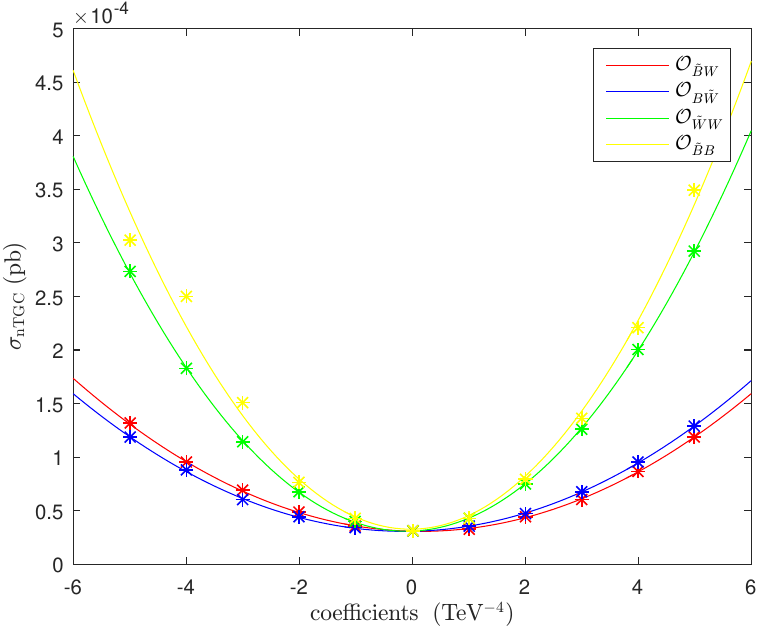}
 		\caption{\label{Fig:fit4}Same as Fig.~\ref{vja1234} but for $e^-p \to  \nu_{\ell}\ell^- \ell^+ j$.}}
 \end{figure}
By scanning the values of coefficients in the region $[-5, 5]\;{\rm TeV}^{-4}$, and the fittings of the cross sections after cut are shown in Fig.~\ref{Fig:fit4}.
The projected sensitivities for $f_{\tilde{B}W}$, $f_{B\tilde{W}}$, $f_{\tilde{W}W}$ and $f_{\tilde{B}B}$ at 2$\sigma$, 3$\sigma$ and 5$\sigma$ are shown at the end of this section in Table~\ref{Tab:constraints}.

\subsection{\label{sec3.5}The process \texorpdfstring{$e^-p \to e^-jjj$}{e-p to e- j j j}}

\begin{figure}[!htbp]
	\centering
	\includegraphics[width=0.45\hsize]{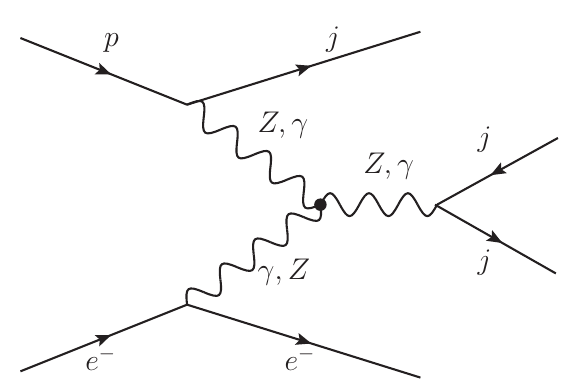}
	\includegraphics[width=0.45\hsize]{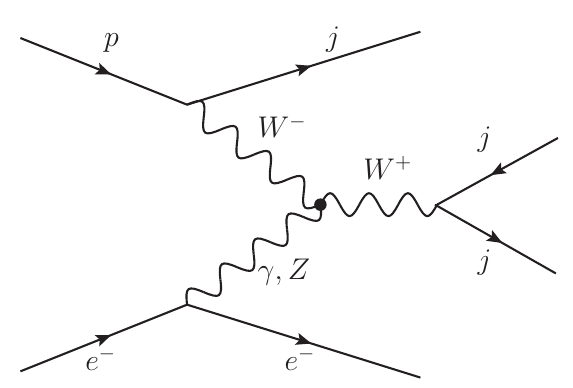}\\
	\includegraphics[width=0.48\hsize]{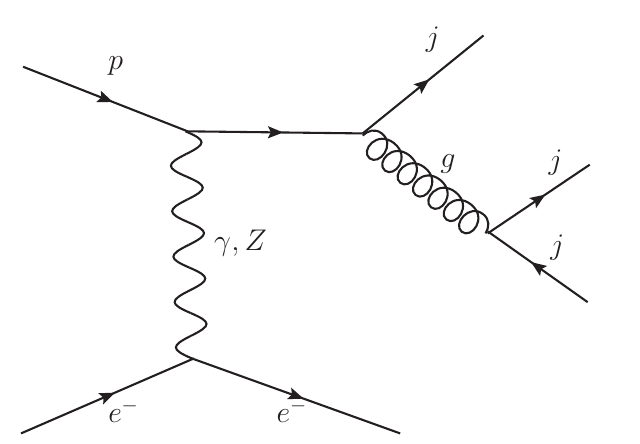}
	\includegraphics[width=0.48\hsize]{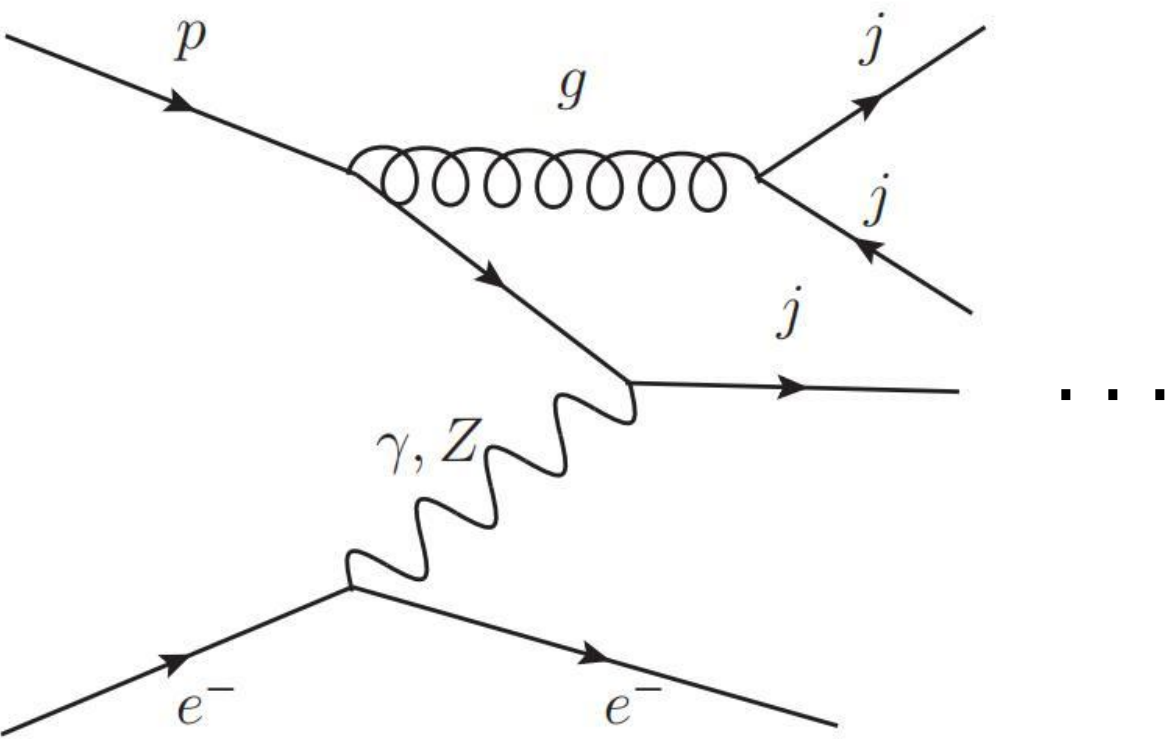}
	\caption{\label{fig:epejjj}Typical Feynman diagrams of nTGCs contribution~(the first row) and the SM backgrounds~(the second row) for the process $e^-p \to e^- jjj$.}
\end{figure}
The $e^-jjj$ signal induced by nTGCs consists of two topological structures induced by five electroweak vertices $ZZZ$, $Z \gamma \gamma$, $ZZ \gamma$, $WW Z$, and $ WW \gamma$. 
The contributions of nTGCs and typical background Feynman diagrams as shown in Fig.~\ref{fig:epejjj}. For the $e^-jjj$ channel, the signal is characterized by three or more jets.
For non-leptonic decay of an energetic $Z$ boson, the jets are collinear and therefore difficult to be separated~\cite{Fu:2021mub,Yang:2021pcf}, as a result, the final state is required to satisfy $N_{jet}\geq 2$, and $N_{\ell}\geq 1$.

\begin{figure}[!htbp]
	\centering{
		\includegraphics[width=0.45\hsize]{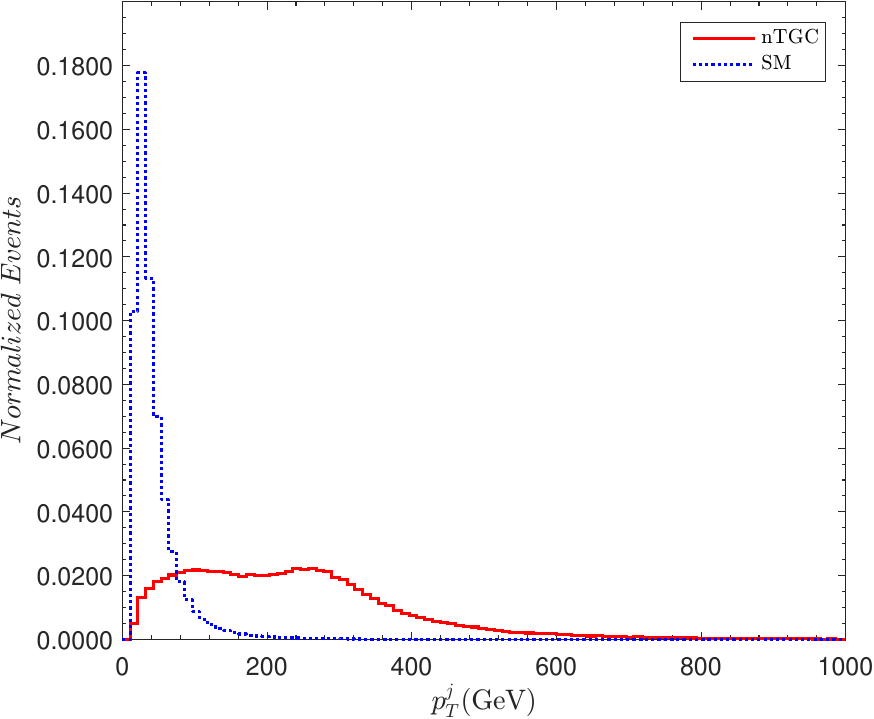}
		\includegraphics[width=0.45\hsize]{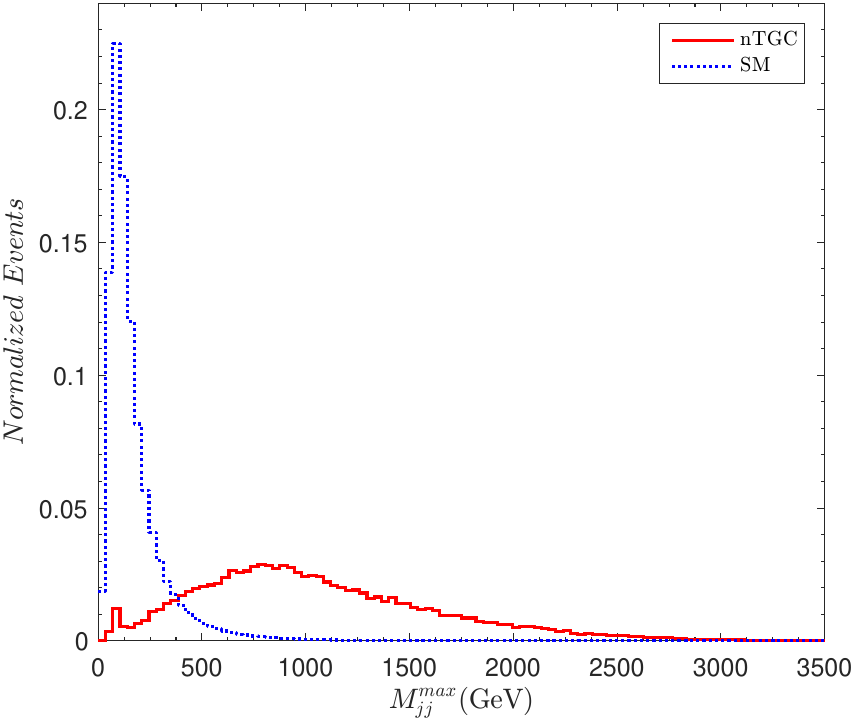}
		\caption{\label{Fig:cut5}The normalized distributions of $p^j_T$ and $M^{max}_{jj}$ for $e^-p \to  e^-jjj$.}}
\end{figure}
When the $W$ or $Z$ boson is energetic, the radiative jets are more energetic, which leads to a large transverse momentum of hardest jet $p_T^{j}$ for the signal. 
In addition, for the background, jets usually radiate from the proton beam, resulting in a smaller invariant mass $M_{jj}$.  
We choose the largest $M_{jj}$ of all pairs of jets~(denoted as $M^{max}_{jj}$), for the signal, if the residue jet is captured, it can be expected $M^{max}_{jj}$ is large.
The normalized distributions of $p^j_T$ and $M^{max}_{jj}$ are shown in Fig.~\ref{Fig:cut5}. 
We apply for cut to $p_T^{j}\geq 300\;\rm{GeV}$ and $M^{max}_{jj} \geq 700\;\rm{GeV}$.

\begin{table}[!htbp]
	\begin{center}
		\begin{tabular}{c|c|c}
			\hline
			& Signal (pb) & Background (pb)  \\
			\hline
			$N_{jet}\geq 2$, $N_{\ell}\geq 1$	& 0.149 	&  0.482   \\
			\hline
			$p_T^{j}\geq 300\;\rm{GeV}$	& 0.054   	& 0.0017  \\
			\hline
			$M^{max}_{jj} \geq 700\;\rm{GeV}$	&   0.038 	&0.00034  \\
			\hline
		\end{tabular}
	\end{center}
	\caption{\label{Tab:cross5}After different cuts, for the $e^-p \to e^-jjj$ process, the cross sections for the signal and SM background at $5.29$ TeV FCC-he with $\mathcal{L} = 2\;\rm{ab}^{-1}$.}
\end{table}
\begin{figure}[!htbp]
	\centering{
		\includegraphics[width=0.6\hsize]{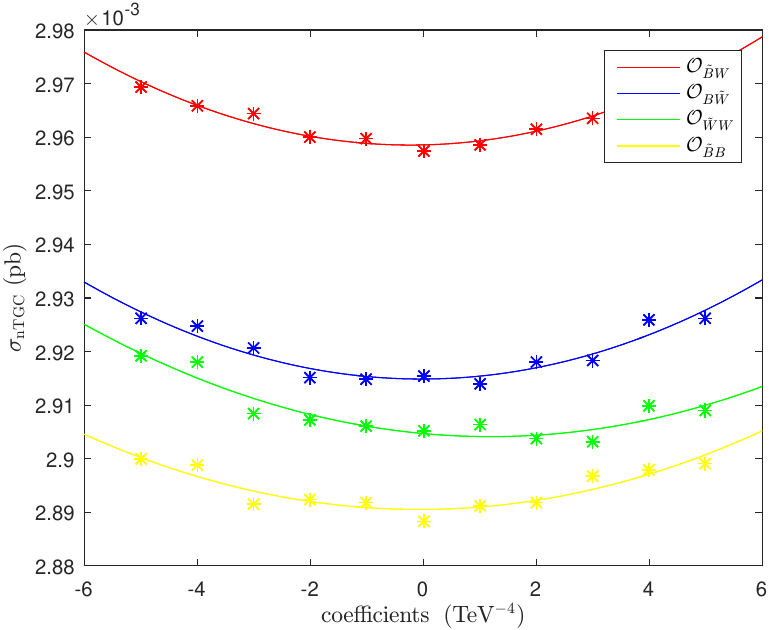}
		\caption{\label{Fig:fit5}Same as Fig.~\ref{vja1234} but for $e^-p \to   e^-jjj$.}}
\end{figure}
The event selection strategy and cut flow are shown in Table~\ref{Tab:cross5}. 
The fittings of the cross sections after cuts are shown in Fig.~\ref{Fig:fit5}.
The expected constraints on the coefficients are obtained and listed at the end of this section in Table~\ref{Tab:constraints}.

\subsection{\label{sec3.6}The process \texorpdfstring{$e^-p \to \nu_e jjj$}{e-p to v j j j}}

\begin{figure}[!htbp]
	\centering
	\includegraphics[width=0.4\hsize]{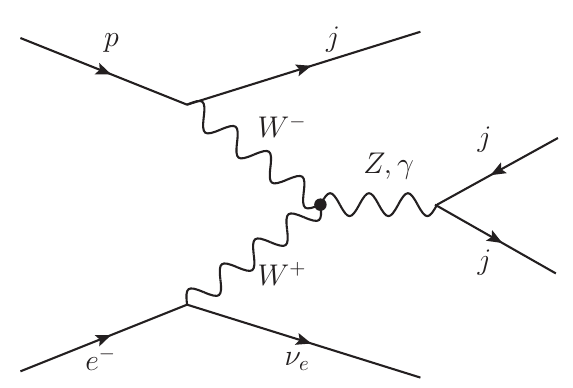}
	\includegraphics[width=0.4\hsize]{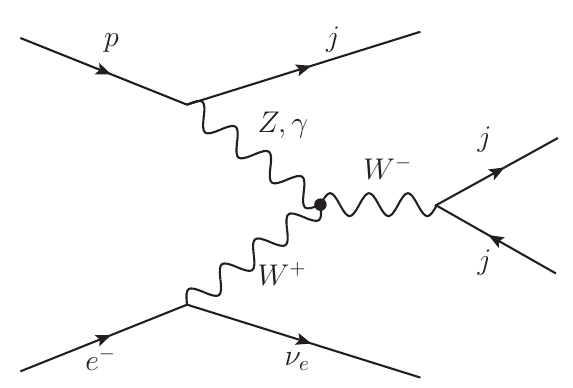}\\
	\includegraphics[width=0.43\hsize]{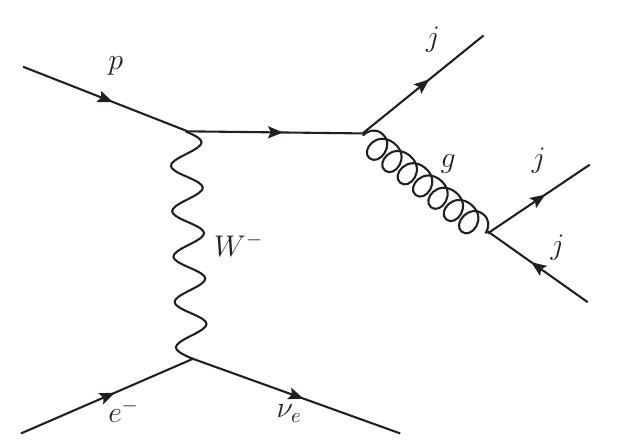}
	\includegraphics[width=0.43\hsize]{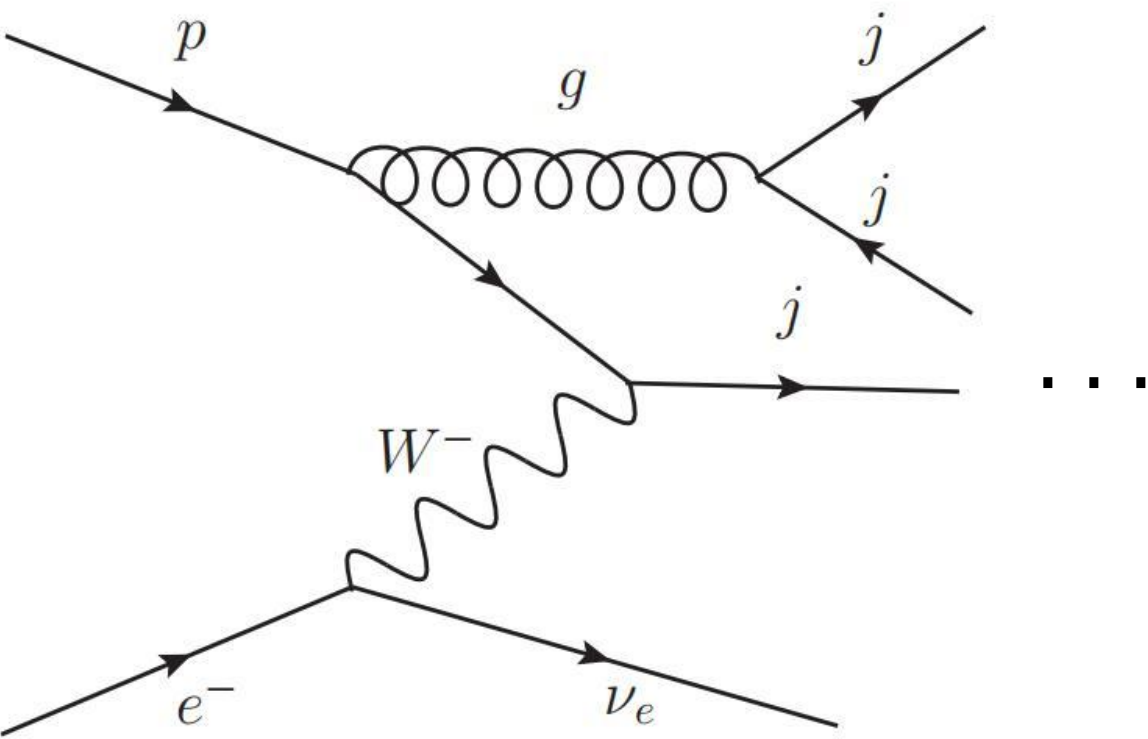}
	\caption{\label{fig:epvjjj}Typical Feynman diagrams of nTGCs contribution~(the first row) and the SM backgrounds~(the second row) for the process $e^-p\to \nu _e jjj$.}
\end{figure}

The typical Feynman diagrams for the process $e^-p \to \nu_e jjj$ are shown in Fig.~\ref{fig:epvjjj}. We required a minimum of two jets in each event, as previously discussed.

The kinematic features are similar to the study in the previous subsection, and $p_T^{j}$ and $M^{max}_{jj}$ are again selected.
Similar to the Fig.~\ref{Fig:cut5}, this subsection is not displayed. In order to reduce background and enhance statistical significance, the cuts are chosen as $p_T^{j}\geq 400\;\rm{GeV}$ and $M^{max}_{jj} \geq 850\;\rm{GeV}$.

\begin{table}[!htbp]
	\begin{center}
		\begin{tabular}{c|c|c}
			\hline
			& Signal (pb) & Background (pb)  \\
			\hline
				$N_{jet}\geq 2$	& 0.146  	&  0.419   \\
			\hline
			$p_T^{j}\geq 400\;\rm{GeV}$	& 0.047   	& 0.0037  \\
			\hline
			$M^{max}_{jj} \geq 850\;\rm{GeV}$	&   0.034 	&0.00058  \\
			\hline
		\end{tabular}
	\end{center}
	\caption{\label{Tab:cross6}For the $e^-p \to \nu_e jjj$ process, the production cross sections of the signal and SM background after the improved cuts applied at $5.29$ TeV FCC-he with $\mathcal{L} = 2\;\rm{ab}^{-1}$.}
\end{table}
\begin{figure}[!htbp]
	\centering{
		\includegraphics[width=0.6\hsize]{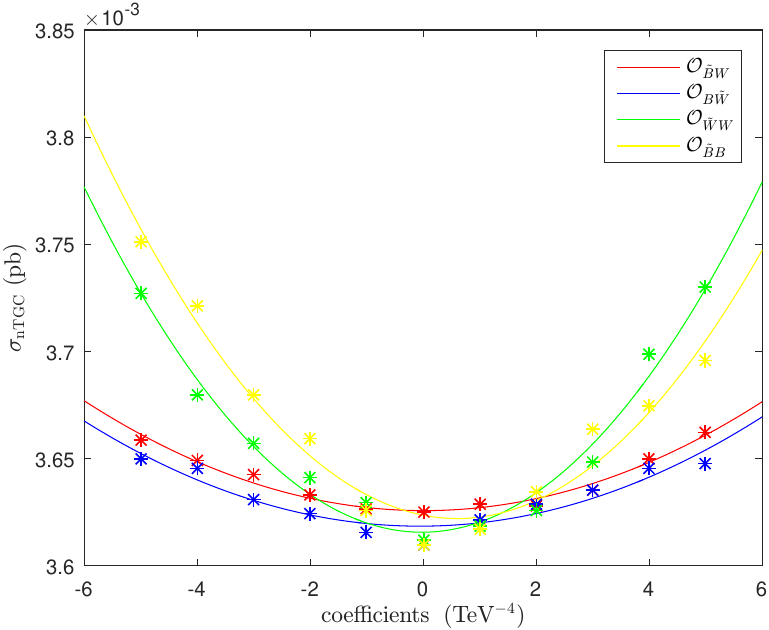}
		\caption{\label{Fig:fit6}Same as Fig.~\ref{vja1234} but for $e^-p \to   \nu_e jjj$.}}
\end{figure}
In Table~\ref{Tab:cross6}, we summarized the cross sections of the signal and background after imposing improved cuts at $5.29$ TeV FCC-he with $\mathcal{L} = 2\;\rm{ab}^{-1}$.
The fittings of the cross sections after cuts are shown in Fig.~\ref{Fig:fit6}.
The expected constraints on the coefficients are obtained and listed in the next subsection in Table~\ref{Tab:constraints}.

\subsection{\label{sec3.7}The combined constraint}

\begin{table*}[!htbp]
\begin{center}
			\begin{tabular}{c|c|c|c|c|c}
				\hline
				process &$\mathcal{S}_{stat}$ & $f_{\tilde{B}W}$ & $f_{B\tilde{W}}$  & $f_{\tilde{W}W}$ & $f_{\tilde{B}B}$  \\
				\hline
				\multirow{4}{*}{$e^-p \to \nu_{e} \gamma j$}	& 2	& $[-1.32, 1.64]$   	& $[-1.64, 1.34]$ & $[-3.02, 2.40]$ & $[-3.88, 4.69]$ \\ \cline{2-6} &  3   &  $[-1.65, 1.98]$&  $[-1.98, 1.68]$& $[-3.63, 3.01]$& $[-4.85, 5.66]$ \\ \cline{2-6} &  5   &  $[-2.20, 2.52]$ &$[-2.53, 2.23]$ &$[-4.63, 4.01]$& $[-6.39, 7.20]$ \\
				\hline
				\multirow{4}{*}{$e^-p \to e^- \gamma j$}	& 2	& $[-2.00, 2.64]$ 	&  $[-2.31, 3.16]$ &  $[-3.72, 4.01]$ & $[-5.55, 4.98]$\\ \cline{2-6} &  3   &  $[-2.85, 2.99]$&  $[-3.08, 3.77]$&  $[-4.70, 5.23]$ &$[-6.59, 6.01]$ \\ \cline{2-6} &  5   &  $[-4.12, 4.33]$ & $[-4.11, 4.76]$ & $[-5.88, 6.86]$  & $[-8.22, 8.00]$  \\
				\hline
				\multirow{4}{*}{$e^-p \to  \ell^- \nu_{\ell}\bar{\nu_\ell}j$}& 2	& $[-1.65, 1.50]$   	& $[-1.37, 1.65]$ & $[-0.89, 1.35]$ &$[-0.75, 1.04]$\\ \cline{2-6} &  3   &  $[-2.00, 1.86]$&  $[-1.71, 1.99]$& $[-1.15, 1.59]$&$[-0.95, 1.24]$ \\ \cline{2-6} &  5   &  $[-2.57, 2.43]$ &$[-2.25, 2.54]$ &$[-1.54, 1.99]$&$[-1.27, 1.56]$ \\
				\hline
				\multirow{4}{*}{$e^-p \to   \nu_{\ell}\ell^- \ell^+ j$}	& 2	& $[-1.39, 1.71]$   	& $[-1.69, 1.42]$ & $[-1.05, 0.85]$&$[-0.90, 0.84]$ \\ \cline{2-6} &  3   &  $[-1.79, 2.11]$&  $[-2.09, 1.82]$& $[-1.30, 1.10]$ &$[-1.13, 1.07]$\\ \cline{2-6} &  5   &  $[-2.52, 2.83]$ &$[-2.82, 2.55]$ &$[-1.74, 1.54]$ &$[-1.54, 1.48]$\\
				\hline
				\multirow{4}{*}{$e^-p \to  e^-jjj$}	& 2	& $[-10.51, 10.05]$   	& $[-10.13, 9.68]$ & $[-10.00, 9.87]$&$[-9.02, 8.74]$ \\ \cline{2-6} &  3   &  $[-12.88, 12.42]$&  $[-12.39, 11.95]$& $[-12.28, 12.15]$&$[-11.68, 11.39]$ \\ \cline{2-6} &  5   &  $[-16.71, 16.25]$ &$[-16.03,  15.59]$ &$[-15.94, 15.80]$&$[-14.37, 14.09]$ \\
				\hline
				\multirow{4}{*}{$e^-p \to   \nu_e jjj$}	& 2	& $[-7.95, 7.35]$   	& $[-7.64, 7.44]$ & $[-4.54, 5.59]$&$[-3.91, 5.12]$ \\ \cline{2-6} &  3   &  $[-9.70, 9.09]$&  $[-9.36, 9.17]$& $[-5.68, 6.74]$ &$[-4.92, 6.14]$\\ \cline{2-6} &  5   &  $[-12.50, 11.89]$ &$[-12.13, 11.93]$ &$[-7.53, 8.58]$&$[-6.56, 7.78]$ \\
				\hline

				\multirow{4}{*}{Combined} & 
2 & $[-1.04, 1.20]$  	& $[-1.16, 1.06]$ &$[-0.80, 0.80]$ & $[-0.66, 0.78]$\\ \cline{2-6} &  
3 & $[-1.31, 1.47]$   	& $[-1.42, 1.35]$ & $[-1.01, 1.03]$ & $[-0.85, 0.94]$ \\ \cline{2-6} &  
5 & $[-1.73, 1.90]$  	& $[-1.84, 1.76]$ & $[-1.36, 1.38]$ & $[-1.10, 1.26]$ \\
				\hline                
			\end{tabular}
\end{center}
\caption{\label{Tab:constraints}The expected constraints on $f_{\tilde{B}W}, f_{B\tilde{W}}, f_{\tilde{W}W}$ and $f_{\tilde{B}B}$ (${\rm TeV}^{-4}$) at FCC-he with $\sqrt{s}=5.29$ TeV and $\mathcal{L} = 2\;\rm{ab}^{-1}$.}
\end{table*}
A summary of the expected constraints from the six processes are shown in Table~\ref{Tab:constraints}. When considering the contribution of nTGCs, we consider a cross-section of the physical process that grows with increasing center-of-mass energy. At sufficiently high energies, this leads to a violation of unitarity, which indicates that SMEFT is no longer valid as a perturbative framework. Unitarity is often used as a criterion for assessing the validity of SMEFT~\cite{Corbett:2014ora,Chen:2023bhu,Remmen:2020uze,Garcia-Garcia:2019oig}. Refs.~\cite{Xie:2025izk,Fu:2021mub} provide unitary constraints for process $ff \to VV$. The process considered in this paper is $VV \to V$, which is somewhat unique and has not been addressed in previous studies. However, based on previous research, it appears that the unitary constraints do not impose particularly stringent restrictions on the expected values of the nTGC coefficients.

To ensure gauge invariance, the $WWZ/WW\gamma$ vertices inevitably appear in the nTGCs operators. 
However, as can be seen from Table~\ref{Tab:channels}, for the VBS processes, different vertices contribute to different channels. 
For instance, process $e^-p\to \nu _e \gamma j$ receives contributions only from $WW\gamma$, while process $e^-p\to e^-\gamma j$ receives contributions only from $Z\gamma\gamma/ZZ\gamma$. 
Therefore, by examining the expected coefficient constraints from them, it can be observed that their sensitivities to the nTGCs are of the same order of magnitude, with process $e^-p\to \nu _e \gamma j$ being slightly more sensitive. 
We believe this is because the initial state of the VBS subprocess in process $e^-p\to \nu _e \gamma j$ involves $W$ bosons, whereas in process $e^-p\to e^-\gamma j$, the VBS initial state involves $Z/\gamma$. 
Since $W$ and $Z$ bosons can have transverse polarizations, the luminosity of $W$ and $Z$ bosons emitted from the beams is logarithmically enhanced at high energies, while photons do not benefit from this luminosity enhancement.

\begin{figure*}[!htbp]
\centering{
\includegraphics[width=0.8\textwidth]{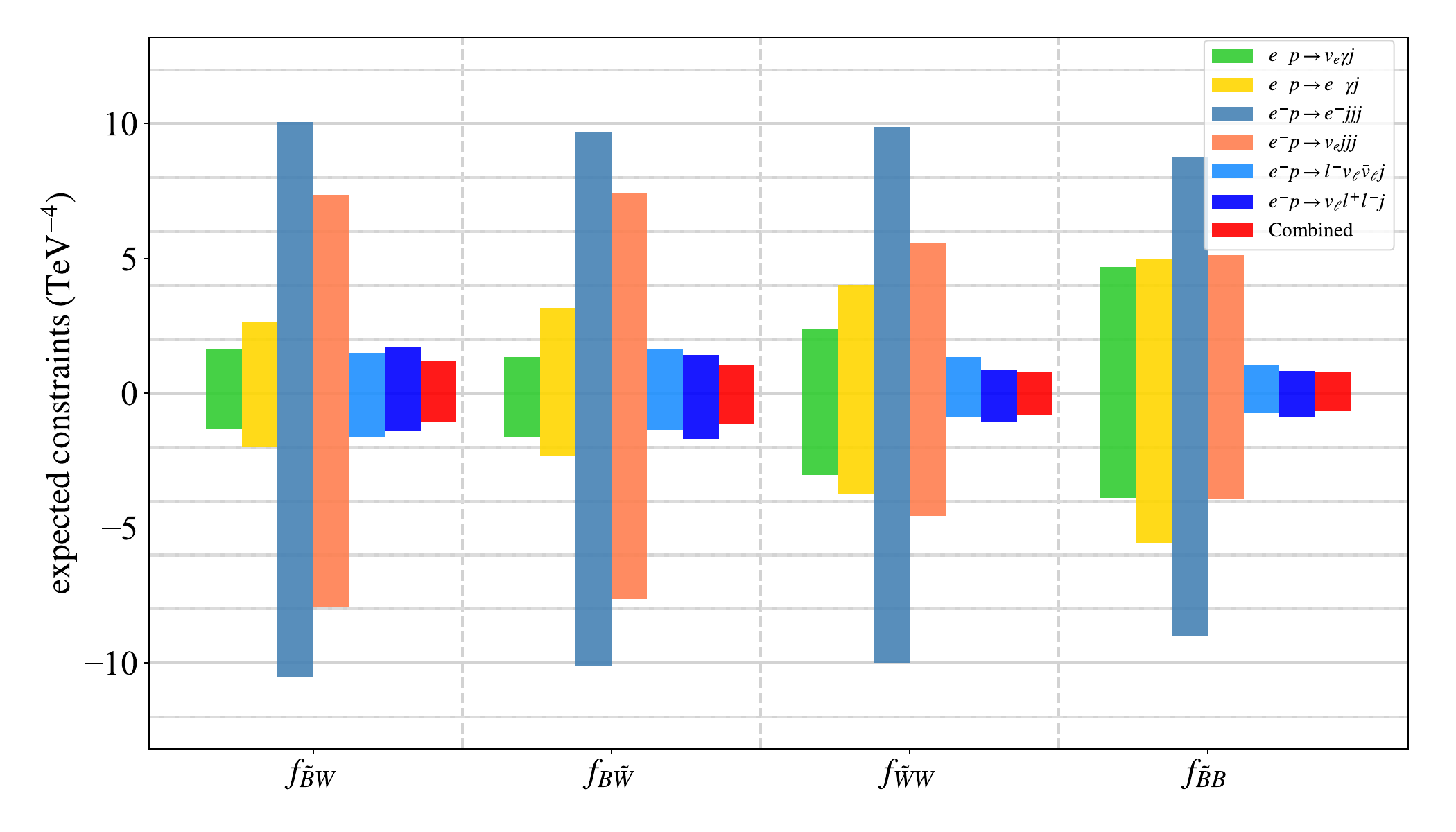}
\caption{\label{Fig:combine}For $\mathcal{S}_{stat}$ equaling $2$, the constraints on $f_{\tilde{B}W}$, $f_{B\tilde{W}}$, $f_{\tilde{W}W}$ and $f_{\tilde{B}B}$ (${\rm TeV}^{-4}$) with combined results.}}
\end{figure*}

By combining the these six processes, tighter constraints can be expected, with the combined signal significance defined as,
\begin{equation}
\begin{split}
\mathcal{S}^{\rm{combine}}_{stat} = \sqrt{\sum{\mathcal{S}^{2}_{stat}}(i)}
\end{split}
\label{eq.5}
\end{equation}
where $S_{stat}(i)$ is the signal significance for each process. 
Combining the data from the six decay modes, the numerical results of the expected constraints for the $f_{\tilde{B}W}$, $f_{B\tilde{W}}$, $f_{\tilde{W}W}$ and $f_{\tilde{B}B}$ coefficients are shown in Table~\ref{Tab:constraints}. For $\mathcal{S}_{stat}=2$, constraints on coefficients for different processes with combined results are shown in Fig.~\ref{Fig:combine}. 

The ATLAS Collaboration has recently performed new measurements of nTGCs at LHC with $\sqrt{s}=13$ TeV and $\mathcal{L} = 140\;\rm{fb}^{-1}$~\cite{ATLAS:2025ply}. The constraints on the coefficients of $f_{\tilde{B}W}$, $f_{B\tilde{W}}$, $f_{\tilde{W}W}$ and $f_{\tilde{B}B}$ at $95\%$ confidence level~(C.L.) are $[-0.54,0.53]$, $[-0.87, 0.95]$, $[-1.90, 1.78]$ and $[-0.37, 0.37]$ (${\rm TeV}^{-4}$), respectively. For $\mathcal{S}_{stat}=2$, it can be observed that the expected constraints on the four operators at $5.29$ TeV FCC-he with $\mathcal{L} = 2\;\rm{ab^{-1}}$ are similar to the LHC results. In addition, Ref.~\cite{Guo:2024qyx} calculates the combined constraints on $f_{\tilde{B}W}$~(${\rm TeV}^{-4}$) for five signal channels of $ZZ$ production at future $e^+ e^-$ colliders with $\sqrt{s}= 250$ GeV, $1$ TeV, and $3$ TeV, corresponding to CEPC, ILC and CLIC, which are $[-2.43, 2.64]$, $[-0.14, 0.36]$ and $[-0.0079, 0.0093]$ at $95\%$ CL, respectively.
For the coefficient $f_{\tilde{B}W}$, the combined constraints at $5.29$ TeV FCC-he are more competitive than those from CEPC experiment. However, compared to ILC results, the combined constraints are slightly less sensitive to nTGCs, which complementing existing searches for nTGCs.

\begin{figure}[!htbp]
	\begin{center}
		\subfigure[$e^-p \to \nu _{e} \gamma j$]{\includegraphics [scale=0.51] {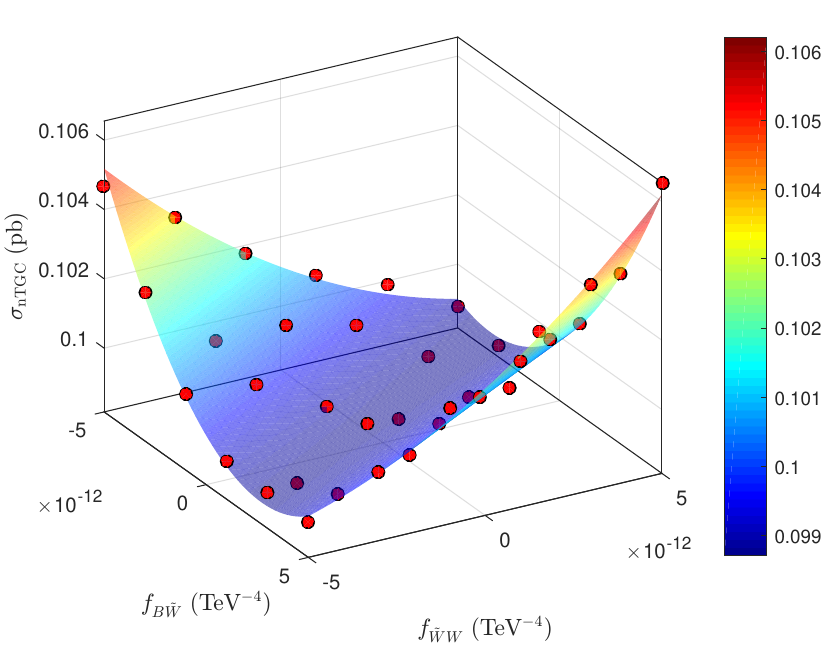}}
		\hspace{0.3in}
		\subfigure[$e^-p \to e^- \gamma j$]{\includegraphics [scale=0.5] {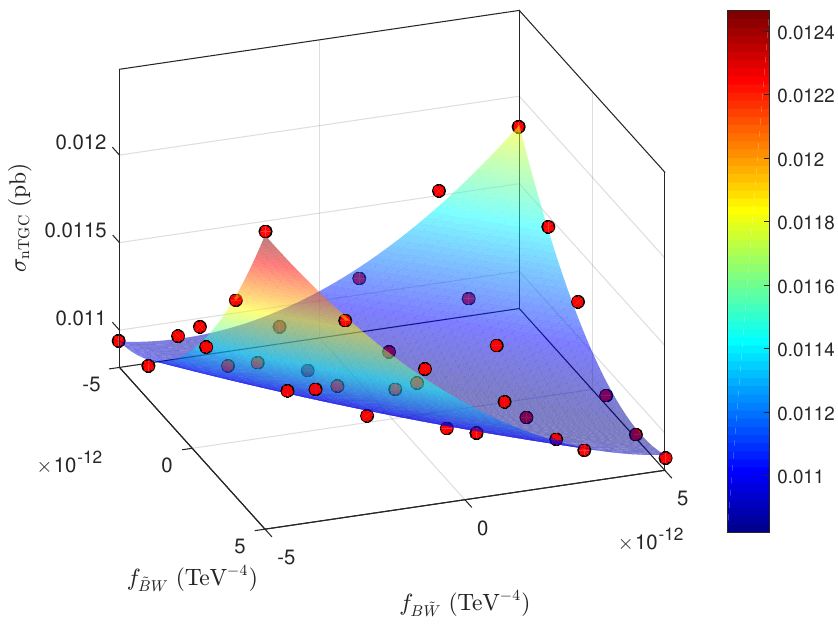}}
		\hspace{0.3in}
		\subfigure[$e^-p \to \ell^- \nu_{\ell}\bar{\nu_\ell}j$]{\includegraphics [scale=0.51] {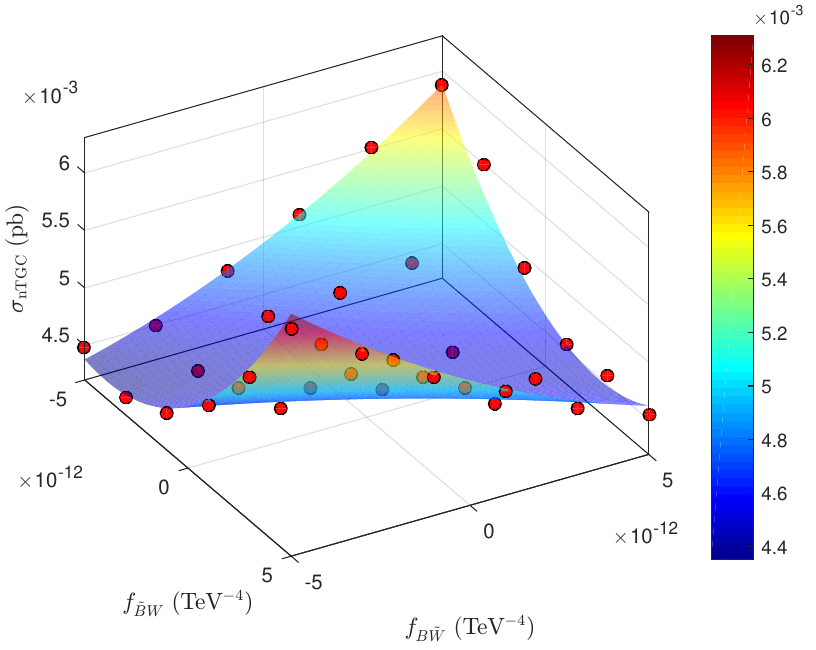}}
		\hspace{0.3in}
		\subfigure[$e^-p \to \nu_{\ell}\ell^- \ell^+ j$]{\includegraphics [scale=0.5] {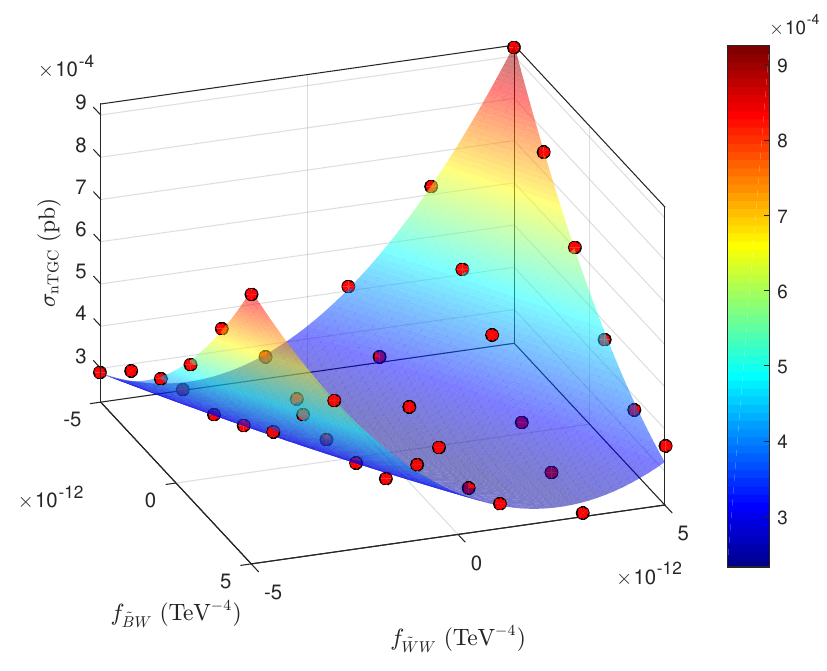}}
		\caption{Randomly select three-dimensional function graphs from the six processes that fit the total cross-section after event selection strategy has been applied to the two operator coefficients.}
		\label{Fig:combinations}
	\end{center}
\end{figure}

The cross section in presence of two operator cofficients~(${f}_{\rm x}$, ${f}_{\rm y}$) can be parameterized as,
\begin{equation}
	\begin{split}
		&\sigma_{\rm{nTGC}} =\sigma_{\rm{SM}} + f_{\rm(x,y,xy)} \times \widehat{\sigma}_{\rm{int}} + f^2_{\rm (x,y)} \times \widehat{\sigma}_{\rm{np}}.
	\end{split}
	\label{eq.3.1.5}
\end{equation}
Where $\widehat{\sigma}_{\rm{np}}$ is NP contribution after cuts to be fitted and $\widehat{\sigma}_{\rm{int}}$ is the interference parameter to be fitted. The coefficients $ f_x, f_y \in \{f_{\tilde{B}W}, f_{B\tilde{W}}, f_{\tilde{W}W}, f_{\tilde{B}B}\}$.
As shown in Eq.~(\ref{eq.3.1.5}), the total cross sections after event selection strategy can be fitted as a quadratic function of the two operator coefficients. Fig.~\ref{Fig:combinations} illustrates 3D plots for several selected cases among the 36 possible combinations of the two operators for the six processes studied in this paper.

\begin{figure}[!htbp]
	\begin{center}
		\subfigure[]{\includegraphics [scale=0.5] {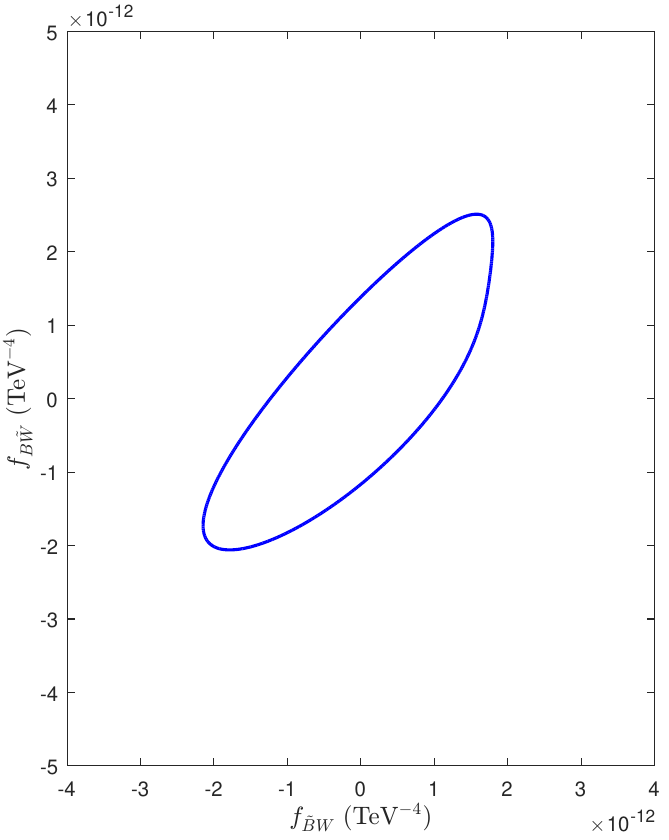}}
		\hspace{0.3in}
		\subfigure[]{\includegraphics [scale=0.5] {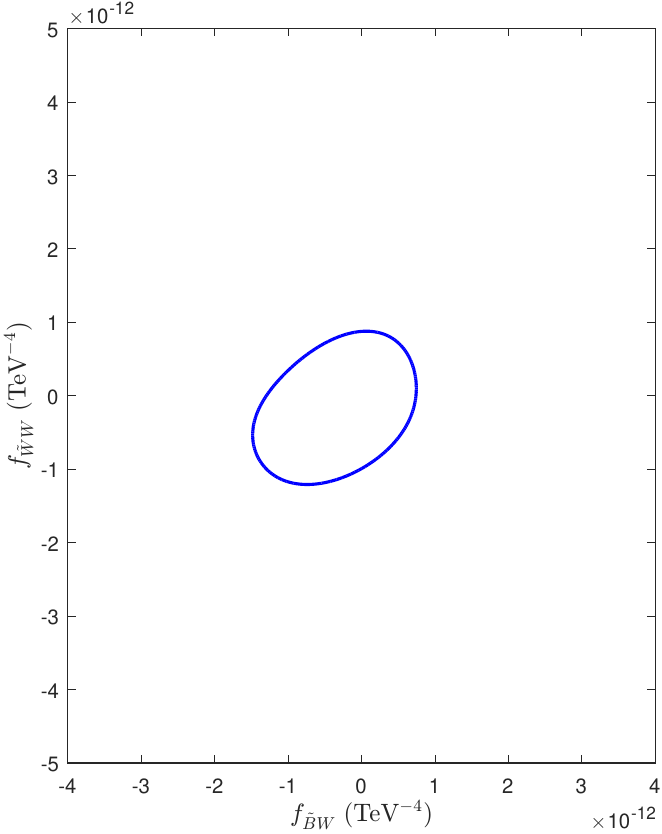}}
		\hspace{0.3in}
		\subfigure[]{\includegraphics [scale=0.5] {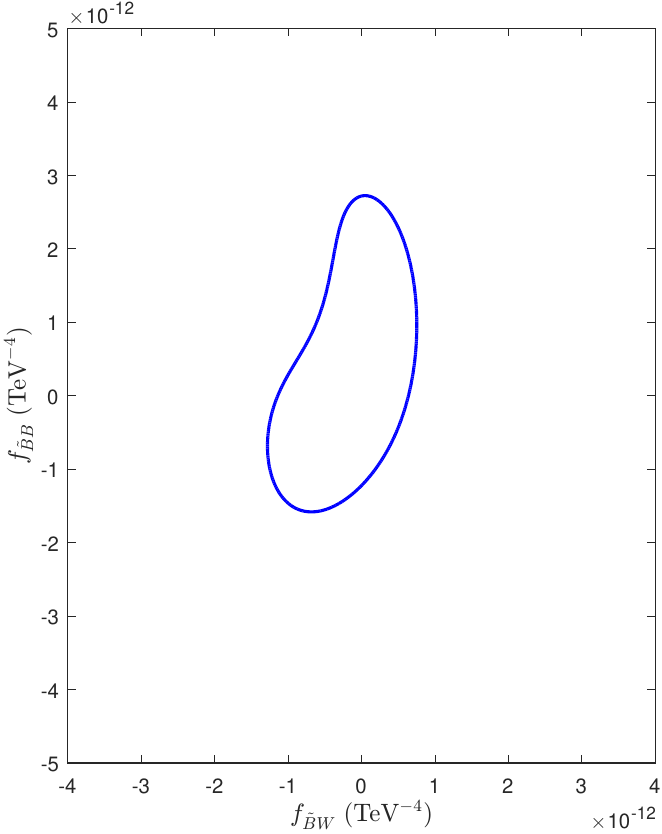}}
		\hspace{0.3in}
		\subfigure[]{\includegraphics [scale=0.5] {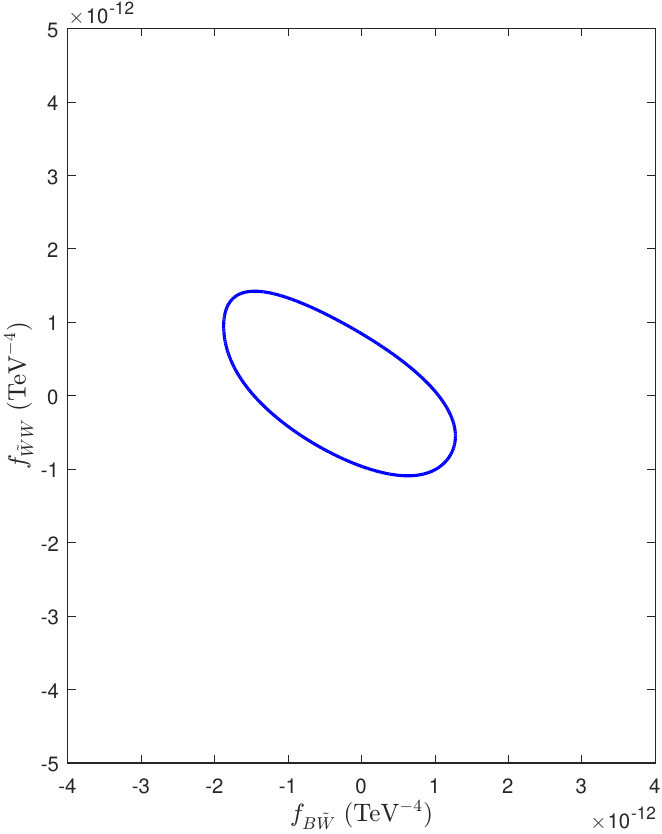}}
			\hspace{0.3in}
		\subfigure[]{\includegraphics [scale=0.51] {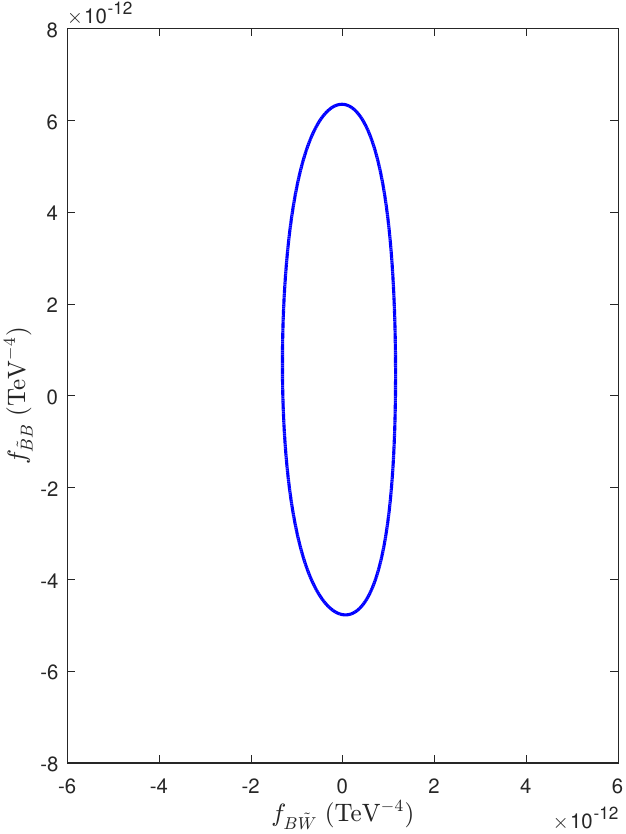}}
			\hspace{0.3in}
		\subfigure[]{\includegraphics [scale=0.51] {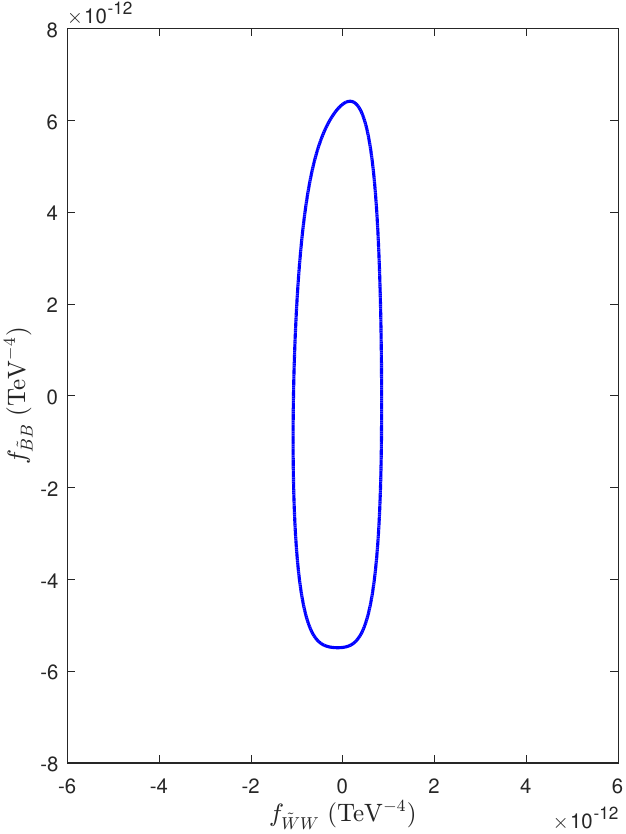}}
		\caption{Two dimensional marginalized projections of coefficients at $95\%$ CL, combining data from six processes.}
		\label{Fig:six}
	\end{center}
\end{figure}

To comprehensively comprehend the effects of interference between operators, we analyzed the two-dimensional marginal projections of the operators at FCC-he with $\sqrt{s}=5.29$ TeV and $\mathcal{L} = 2\;\rm{ab}^{-1}$.
Using Eq.~\ref{eq.5}, we obtain two dimensional marginalized projections of the function of any two operator coefficients~${f}_{\rm x}$, ${f}_{\rm y}$~(${\rm TeV}^{-4}$) by combining data from six decay modes at $95\%$ C.L  as shown in Fig.~\ref{Fig:six}. It can be seen that the Fig.~\ref{Fig:six} are not ellipses, because the superposition of various interference effects from different channels in a non-trivial pattern.

In this work, we present constraint results for both single-operator and two-operator models. The single-operator results~(Table~\ref{Tab:constraints}) can be used for comparison with data from existing studies (e.g., LHC, CEPC, ILC), as they provide a direct, reproducible reference benchmark that facilitates cross-comparisons between different studies.

However, in actual new physics scenarios, multiple operators may coexist and produce interference effects. To evaluate this possibility, we further performed a two-parameter fit, yielding two dimensional marginalized projections~(Fig.~\ref{Fig:six}). These results clearly demonstrate the correlation between different operator coefficients, when two operators are allowed to vary together, the constraint region tends to stretch or tilt due to the contribution of interference terms. The different decay channels exhibit distinct interference effects, and the superimposed results are non-trivial. Compared to single-operator constraints, two-dimensional combined constraints are generally more conservative. This is because the single-operator method ignores the possibility that other operators may partially cancel out signal effects through interference.

\section{\label{level4}Summary }

Currently, nTGCs provide an important window for probing NP beyond the SM, which can be used to directly probe new physics by dimension-8 SMEFT operators. 
In the framework of the SMEFT, the expected constraints on the coefficients of $f_{\tilde{B}W}$, $f_{B\tilde{W}}$, $f_{\tilde{W}W}$ and $f_{\tilde{B}B}$ are investigated at FCC-he with $\sqrt{s}=5.29$ TeV and $\mathcal{L} = 2\;\rm{ab}^{-1}$ though the processes $e^-p \to \nu_{e} \gamma j$, $e^-p \to e^- \gamma j$, $e^-p \to \ell^- \nu_{\ell}\bar{\nu_\ell}j$, $e^-p \to \nu_{\ell}\ell^- \ell^+ j$, $e^-p \to e^-jjj$ and $e^-p \to \nu_e jjj$, respectively.
For the $f_{\tilde{B}W}$ coefficient, the greatest sensitivity is typically provided by the $e^-p \to \nu_{e} \gamma j$ process, while for the $f_{B\tilde{W}}$, $f_{\tilde{W}W}$ and $f_{\tilde{B}B}$ coefficients, it is the $e^-p \to \ell^- \nu_{\ell}\bar{\nu}_{\ell}j$ process. 
In addition, the combining results from multiple channels can further increase the sensitivity to nTGCs.

It should be noted that the operator set in Eq.(2) is incomplete. While a more complete set of nTGC operators now exists, Eq.(2) was proposed earlier. In early works, researchers primarily used the operators from Eq.(2) for studies, and we compare our results with theirs.
The purpose of this paper is to investigate whether the $e^-p$ collider is sensitive to nTGCs. Generally, if a collider is sensitive to these operators, it is sensitive to this whole group.
The expected constraints on $f_{\tilde{B}W}$ , $f_{B\tilde{W}}$, $f_{\tilde{W}W}$ and $f_{\tilde{B}B}$ are compared with ones obtained at $13$ TeV LHC and $250$ GeV CEPC. 
The results show that the sensitivity of FCC-he in detecting nTGCs is comparable to that of LHC and provides tighter constraints than the CEPC. 
It is important to emphasize that the results presented in this paper demonstrate the ability of $e^-p$ colliders to detect dimension-8 operators within the SMEFT. The primary advantage of $e^-p$ colliders does not necessarily lie in providing the most stringent constraints, but rather in offering complementary insights into the study of nTGCs through different initial states, a unique experimental environment, and a distinct dynamical coverage, thereby serving as an important cross-validation and supplement to research conducted on $pp$ and $e^+ e^-$ colliders.

\begin{acknowledgements}
This work was supported in part by the National Natural Science Foundation of China under Grants No.~12147214 and No.~12575106, the Basic Research Projects of Universities in Liaoning Province~(Grant No.~LJKMZ20221431), the Department of Education of Liaoning Province~(Grant No.~LJ212510165022) and the Projects of Liaoning Normal University~(Grant No. 2024BSL018).
\end{acknowledgements}
	
\bibliography{ntgc-ep}

\end{document}